\documentclass[submission, Phys]{SciPost}
\usepackage[T1]{fontenc}
\usepackage{booktabs}
\usepackage{graphicx}
\usepackage{xspace}
\usepackage{multirow}
\usepackage{amssymb,amsmath}
\usepackage{mathtools}
\usepackage{cleveref}
\usepackage{xstring}
\usepackage[normalem]{ulem}
\usepackage{algorithm}
\usepackage{algpseudocode}
\usepackage{ulem}
\hypersetup{colorlinks, linkcolor={red!50!black}, citecolor={blue!50!black},
  urlcolor={blue!80!black}}

\begin{document}

\thispagestyle{empty}

%\def\thefootnote{\fnsymbol{footnote}}

%\vspace*{1cm}

\begin{center}

\begin{Large}
\textbf{\textsc{Inferring correlated distributions: boosted top jets}}
\end{Large}

\vspace{1cm}

{\sc
Ezequiel~Alvarez$^{a}$%
\footnote{{\tt \href{mailto:sequi@unsam.edu.ar}{sequi@unsam.edu.ar}}}%
, Manuel~Szewc$^{b,a}$%
\footnote{{\tt \href{mailto:szewcml@ucmail.uc.edu}{szewcml@ucmail.uc.edu}}}%
, Alejandro~Szynkman$^{c}$%
\footnote{{\tt \href{mailto:szynkman@fisica.unlp.edu.ar}{szynkman@fisica.unlp.edu.ar}}}%
, Santiago~Tanco$^{c}$%
\footnote{{\tt \href{mailto:santiago.tanco@fisica.unlp.edu.ar}{santiago.tanco@fisica.unlp.edu.ar}}}%
, Tatiana~Tarutina$^{c}$%
\footnote{{\tt \href{mailto:tarutina@fisica.unlp.edu.ar}{tarutina@fisica.unlp.edu.ar}}}%
}

\vspace*{.7cm}

{\sl
$^a$International Center for Advanced Studies (ICAS) and ICIFI-CONICET, UNSAM, \\
25 de Mayo y Francia, CP1650, San Mart\'{\i}n, Buenos Aires, Argentina

\vspace*{0.1cm}
$^b$Department of Physics, University of Cincinnati, \\
Cincinnati, Ohio 45221,USA
\vspace*{0.1cm}

$^c$IFLP, CONICET - Dpto. de F\'{\i}sica, Universidad Nacional de La Plata, \\ 
C.C. 67, 1900 La Plata, Argentina

\vspace*{0.1cm}

}

\end{center}

\vspace{0.1cm}

%\linenumbers
\begin{abstract}
\noindent 
Improving the understanding of signal and background distributions in signal-region is a valuable key to enhance any analysis in collider physics.  This is usually a difficult task because --among others-- signal and backgrounds are hard to discriminate in signal-region, simulations may reach a limit of reliability if they need to model non-perturbative QCD, and distributions are multi-dimensional and many times may be correlated within each class.  Bayesian density estimation is a technique that leverages prior knowledge and data correlations to effectively extract information from data in signal-region.  In this work we extend previous works on data-driven mixture models for meaningful unsupervised signal extraction in collider physics to incorporate correlations between features.  Using a standard dataset of top and QCD jets, we show how simulators, despite having an expected bias, can be used to inject sufficient inductive nuance into an inference model in terms of priors to then be corrected by data and estimate the true correlated distributions between features within each class. We compare the model with and without correlations to show how the signal extraction is sensitive to their inclusion and we quantify the improvement due to the inclusion of correlations using both supervised and unsupervised metrics.
\vskip 1cm
~
\end{abstract}

\def\thefootnote{\arabic{footnote}}
\setcounter{page}{0}
\setcounter{footnote}{0}

\newpage

\tableofcontents

\section{Introduction}\label{sec:intro}

The Large Hadron Collider (LHC) is entering the High Luminosity era.  However, in many cases the $\mathcal{O}(10)$ gain in luminosity will not translate straightforwardly to a relevant gain in precision unless it is accompanied with a more detailed understanding of the detectors and the underlying physics and distributions followed by the variables involved in each process. In fact, detector knowledge and physics modeling are intimately related, with improvements in one area allowing for improvements in the other.

In this work we are interested in developing techniques to infer underlying distributions corresponding to LHC analyses. From a statistical point of view, a given selection of events can always be modeled as a mixture of classes: each event belongs to a given class, which could be, for example, signal and the corresponding backgrounds~\footnote{Interference between processes implies that they need to be grouped in the same class. In this work we assume the signal does not interfere with the background and thus it is always its own separate class.}.  Bayesian techniques are a powerful means of addressing these kinds of problems. By modeling events as samples from a probabilistic distribution, one can unfold the probabilistic model to latent variables. This unfolding allows for the incorporation of important prior knowledge for each one of the latent variables, thereby enhancing the posterior performance and deepening the understanding of the system.  These techniques have been widely used in the Machine Learning industry (see Ref.~\cite{bw} for a description of the recommended Bayesian workflow) and have been known to the LHC community for quite some time (see {\it e.g.} Ref.~\cite{Cranmer:2014lly} for discussions about mixture models in the context of frequentists and/or Bayesian analyses) with broader advances in Bayesian computation becoming more popular in LHC applications in recent years~\cite{Dillon:2019cqt,Dillon:2020quc,Dillon:2021aeo,Brivio:2022hrb,Yallup:2022rjd,Fowlie:2024dgj,Albert:2024zsh,Fowlie:2025kna,Alvarez:2021hxu,Alvarez:2021zje,Alvarez:2022kfr,Alvarez:2022qoz,Alvarez:2024doi,Alvarez:2024owq}.  

To ensure that inference yields appropriate distributions, a reasonable inductive bias needs to be baked into the model. Focusing on collider applications, there have been different strategies. The more common, and most storied, tradition is to implicitly define the class-dependent distributions using simulated events and consider as free parameters only the class fractions. This is the most powerful approach provided that the simulators are a perfect model of the data. However, we know this not to be the case, and so it is standard practice to extend the model using nuisance parameters to account for different sources of mismodeling, including by the simulators themselves. This is not guaranteed to be sufficient, specially for the high-complexity, high-statistic environment of the LHC, and thus data-driven techniques have long been a staple of the community toolkit\cite{abcd95,abcd97,Bardhan:2024zla}. However, data-driven techniques also need inductive bias, mainly to restrict the possible shapes of the probability distributions involved. For continuous variables, one can achieve this by imposing a particular parametric family of high-dimensional distributions, such as a multivariate gaussian, to which each class-dependent distribution belongs. For discrete variables, such as bin counts, this is trickier to achieve. 

Another possibility that side-steps the need to fully specify a multi-dimensional distribution but still restricts the problem is to assume that all relevant observables --or the dimensions of the problem-- are {\it conditionally independent} or uncorrelated~\footnote{In an abuse of language, but to align with the terminology used in the community, throughout the work we use uncorrelated for independent.} within each class.  This results in a large enhancement in the inferring power, since the problem is reduced to learning sets of one-dimensional distributions and there is no need to add parameters to describe the dependence between the observables and there is no complication coming from the correlation among the characterizing observables.  Examples of previous works exploiting conditional independence can be found in Refs.~\cite{4tops,Metodiev:2017vrx,PhysRevD.101.095004,PhysRevD.104.035003,Kasieczka:2020pil,Klein:2025lbj}.  One should note that, even when conditional independence may not be exact, it may hold up approximately and, for finite data, can be a valid working assumption. However, the estimators will not be unbiased and will not converge to the class fractions in the limit of infinite data (as seen in {\it e.g.} Ref.~\cite{4tops}). Explicit attempts to incorporate conditional dependence rely on theoretical knowledge of factorization properties~\cite{Metodiev:2023izu,Desai:2024yft} or assume weak dependence and rely on the interpolation capabilities of deep neural networks~\cite{Hallin:2021wme,Hallin:2022eoq,Raine:2022hht}.

In this work, we are interested in the study of systems with correlated observables where conditional independence does not hold, but we do not have any explicit parametric shape we can use to restrict the functions and the simulators are imperfect.  As a first step towards a more complete inference, we assume that the correlations between the variables are well described by the simulations, but not their marginalized distributions.  This choice is a compromise between approaches, where the simulator provides the relationship between observables which are used to unfold the data to learn appropriate underlying one-dimensional distributions. Thus, we avoid fully specifying the multi-dimensional structure of the data in terms of a parametric distribution without needing to assume conditional independence. In future works we envisage to also infer the correlation starting from biased priors, while maintaining some structure that allows us to avoid adding too many parameters.

To instantiate the above paradigm we study in this work top-quark related observables, in particular those from hadronically decaying boosted top quark.  In general, these objects are relevant in many searches at the LHC, both for precision physics~\cite{ATLAS:2019kwg,CMS:2020tvq,ATLAS:2021dqb} and to search for beyond the standard model effects~\cite{CMS:2016jce,ATLAS:2022ozf,Araz:2023axv}. Boosted top quarks are observed as fat jets (which we name top jets for convenience) at LHC events, and two relevant characterizing observables of these objects are the number of clusters obtained from re-clustering the constituent particles of the jet with a smaller $R$\cite{Forshaw:1999iv,Catani:1993yx} ($N_{\mathrm{clus}}$) and the jet mass ($\mathrm{Mass}$). The clusters within a fat jet are computed through FastJet \cite{Cacciari:2011ma} and the jet mass is computed using the four-momentum of the constituents of the fat jet.  For a top jet, these two observables are correlated and therefore are a suitable set of variables to test the proposed framework.   The background for hadronically decaying boosted top quarks consists of QCD jets, which originate from a quark or a gluon and have very different cluster and mass distributions.  However, since QCD jets are overwhelmingly more abundant than top jets, they end up being a large background even within a pre-selection of events targeting top jets.  

The objective of this work is to consider a selection of events containing top and QCD jets and infer their distributions in $\vec x = \{N_{\mathrm{clus}}, \mathrm{Mass}\}$ for each class, even if we start the inference from a biased prior. A partial goal of this set up is to work out the possibility that Monte Carlo simulations may not match the data in what respects to the marginalized $\{N_{\mathrm{clus}}$, $\mathrm{Mass}\}$ distributions.  It is a partial goal because, for simplicity, along the work we assume that the correlations in the form of transfer matrices are well described by simulations.  It is worth mentioning that one should expect that simulations at some level do not match the data, since programs such as Pythia \cite{Bierlich:2022pfr}, Herwig \cite{Bahr:2008pv}, etc., need to use phenomenological models to describe non-perturbative QCD effects, and therefore there is an inherent bias that forbids complete match between simulation and data distributions.

This article is organized as follows. In the next section we introduce the statistical models with and without the assumption of conditional independence and discuss the details of their implementation. In section~\ref{sec:results} we perform the inference for both models and different priors; and assess the quality of the obtained results. In section~\ref{sec:conclusions} we summarize the results and present our conclusions. Further details on the estimation of the transfer matrix necessary to account for correlations are given in Appendix~\ref{app:em_algorithm}.

\section{Mixture of multinomial with correlations}\label{sec:mixturemodel}

Mixture models are a crucial tool for science, since often the data in a problem consist of a collection of independently distributed elements, or data points, which are each sampled from one among a set of possible classes.  For many LHC analyses, we are interested in one class in particular, which we label as signal and that we want to characterize in the presence of all other classes, labeled as contributing backgrounds.  Along this work we are interested in studying top events in which a boosted top decays hadronically forming a single jet. Therefore, the dataset is a collection of jets and the classes are top (signal) or QCD (background) jets.

Mixture models allow us to construct probabilistic models of the data, by first selecting an appropriate collection of measurable observables $\vec{x}$ and defining the probability of a single jet as
\begin{eqnarray}
    p(\vec{x}) =\sum_{k=\{0,1\}} \pi_k\, p(\vec{x}|k)\,,
    \label{eq:mixtureModel}
\end{eqnarray}
where $k=0$ corresponds to the QCD background while $k=1$ corresponds to a top jet, $\pi_k$ is the probability of sampling a jet from jet class $k$ and $p(\vec{x}|k)$ is the class-dependent probability of a particular observable value $\vec{x}$. Modeling the class-dependent probability distributions corresponds to capturing the relevant physics.  In particular, if we consider distributions depending on tensors of parameters $\boldsymbol{\theta}^{k}$, $p(\vec{x}|\boldsymbol{\theta}^{k})$, we can obtain the posterior probability distribution over $\boldsymbol{\zeta}=\{(\pi_k, \boldsymbol{\theta}^k), k=0,1\}$ given $N$ measured jets $\boldsymbol{X}=\{\vec{x}_1,\dots,\vec{x}_{N}\}$
\begin{eqnarray}
    p(\boldsymbol{\zeta}|\boldsymbol{X}) = \frac{p(\boldsymbol{X}|\boldsymbol{\zeta}) p(\boldsymbol{\zeta})}{p(\boldsymbol{X})}\,.
\end{eqnarray}
Here $p(\boldsymbol{X}|\boldsymbol{\zeta})=\prod_{n=1}^{N}p(\vec{x}_{n}|\boldsymbol{\zeta})$ for independent jets, $p(\boldsymbol{\zeta})$ is the prior distribution over parameters and $p(\boldsymbol{X})=\int d\boldsymbol{\zeta}\ p(\boldsymbol{X}|\boldsymbol{\zeta}) p(\boldsymbol{\zeta})$ is the evidence or marginal likelihood. The posterior can be estimated via approximate techniques as in Ref.~\cite{Alvarez:2022qoz}. However, for the probabilistic models considered in this work we obtain samples from the posterior by implementing the model in the probabilistic programming language {\tt Stan}~\cite{stan}.

The choice of parametric distributions $p(\vec{x}|\boldsymbol{\theta}^{k})$ depends on the choice of measured variables. In this work we consider the case where the characterizing observable $\vec{x}$ corresponds to the number of clusters in the jet~\footnote{The clustering is performed with the $k_T$ algorithm \cite{Ellis:1993tq} in FastJet setting $R=0.3$ and a cut $p_T > 3$ GeV over the constituents of the fat jet.}  and the mass of the jet, $\vec{x}=\{N_{\mathrm{clus}},\mathrm{Mass}\}$. We bin the mass of the jet in 15 bins between $m_{\mathrm{min}}=150$ GeV and $m_{\mathrm{max}}=225$ GeV, while the $N_{\mathrm{clus}}$ is a naturally discrete variable that in our framework takes 5 possible values. Because the most general distribution of a two-dimensional discrete variable is a multinomial, one might be tempted to describe the class-dependent likelihoods as
\begin{eqnarray}
p(N_{\mathrm{clus}} \in  \mathrm{bin}_{i},\mathrm{Mass} \in \mathrm{bin}_{j}|k)\equiv p(i,j|k) = \theta^{k}_{ij}\,,
\end{eqnarray}
with
\begin{eqnarray}
\sum_{i=1}^{D_1}\sum_{j=1}^{D_2} \theta^{k}_{ij} = 1\,,
\end{eqnarray}
where $D_{1}$ and $D_{2}$ are the number of bins for each variable and $\theta^{k}_{ij}$ the parameters of class $k$. This runs into the problem of attempting to describe a two-dimensional distribution with $D_{1}\times D_{2}-1$ parameters $p(i,j)$ with a mixture of $K$ distributions of the same number of parameters $p(i,j|k)$, which is ill-posed and leads to mode collapse. To avoid over-parameterizing the data, we need to include additional structure in our probabilistic model. This additional structure or model bias allows us to extract useful classes. The {\it arts et m\'{e}tiers} of Bayesian inference is to select the appropriate model for the task at hand, with its compromise between bias and performance. As we show below, a multi-dimensional mixture model is easily addressed under a hypothesis of conditionally independent classes, since the number of parameters in each class gets drastically reduced (in this two dimensional case) from $D_{1}\times D_{2}-1$ to $D_{1} + D_{2} - 2$.  However, if conditional independence does not hold in the problem then correlation matters and the problem becomes considerably more difficult.   Addressing the latter case is the objective of this work.  We describe both cases in the following paragraphs, focusing on the latter, which is the case of interest.

\subsection{Conditionally independent model}\label{sec:fullindependence}

\begin{figure}[h]
    \centering
    \includegraphics[width=0.5\linewidth]{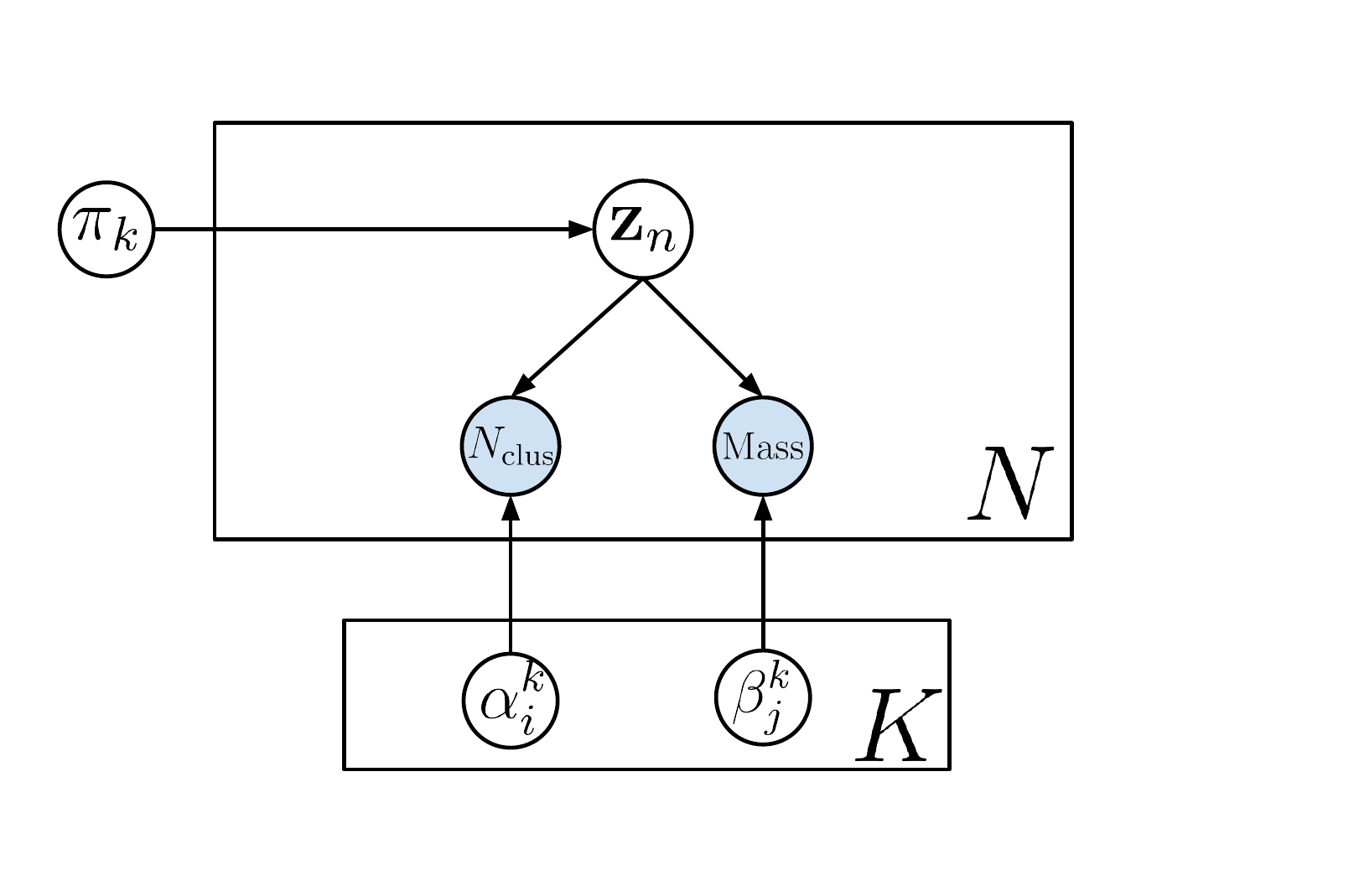}
    \caption{ Graphical model for the conditionally independent model. Solid white (blue) circles represent latent (observed) variables. Boxes represent the repeated sampling of variables.}
    \label{fig:graphical_model_full_independence}
\end{figure}

The simplest model that we can think of is the one that constrains $\theta^{k}_{ij}$ the most by assuming conditional independence.  Following Ref.~\cite{4tops}, we impose per-class conditional independence

\begin{eqnarray}
p(i,j|k) &=& p(i|k)p(j|k)\,,
\end{eqnarray}
or in other words
\begin{eqnarray}
\theta^{k}_{ij} &=& \alpha^{k}_{i}\beta^{k}_{j}\,,
\end{eqnarray}
where $\alpha^{k}_{i}=p(i|k)$ and $\beta^{k}_{j}=p(j|k)$, satisfying $\sum_{i=1}^{D_1}\alpha^{k}_{i}=1$ and $\sum_{j=1}^{D_2}\beta^{k}_{j}=1$ respectively. Here the $\alpha_i^k$ and $\beta_j^k$ are the parameters of the probabilistic multinomial model assumed for the system. The resulting graphical model is shown in Fig.~\ref{fig:graphical_model_full_independence} and the full likelihood can be written as
\begin{eqnarray}
    p(\boldsymbol{X}|\boldsymbol{\zeta})=\prod_{n=1}^{N}\left(\sum_{k=1}^{K}\pi_{k}\alpha^{k}_{i_n}\beta^{k}_{j_n}\right)\,,
\end{eqnarray}
which can be usefully rewritten as
\begin{eqnarray}
    p(\boldsymbol{X}|\boldsymbol{\zeta})=\prod_{i=1}^{D_{1}}\prod_{j=1}^{D_{2}}\left(\sum_{k=1}^{K}\pi_{k}\alpha^{k}_{i}\beta^{k}_{j}\right)^{N_{ij}}\,,
\end{eqnarray}
where $N_{ij}$ is the total number of events populating bin $(i,j)$ and $\sum_{i=1}^{D_{1}}\sum_{j=1}^{D_{2}}N_{ij}=N$.

\subsection{A simulation-assisted model for the lack of conditional independence}\label{sec:simulationdependence}

As shown in Ref.~\cite{4tops}, if the assumption of conditional independence is incorrect, for a sufficiently large number of events the bias of the model introduced in~\cref{sec:fullindependence} will shift the posterior from the true values, because the corresponding hypotheses do not hold. Since in this work we consider a large dataset, this bias is present as shown in section~\ref{sec:results}. To incorporate correlations, we introduce two additional transfer matrices that convolute the truly factorized per-class distributions
\begin{eqnarray}
\theta^{k}_{ij} = \sum_{i'=1}^{D_{1}}\sum_{j'=1}^{D_{2}}C^{k}_{ij;i'j'}\alpha^{k}_{i'}\beta^{k}_{j'}\,,
\end{eqnarray}
with $\sum_{i=1}^{D_{1}}\sum_{j=1}^{D_{2}}C^{k}_{ij;i'j'}=1$ for all $i',j',k$. This corresponds to writing
\begin{eqnarray}\label{eq:transfer_likelihood}
p(ij|k) = \sum_{i'=1}^{D_{1}}\sum_{j'=1}^{D_{2}}p(i,j,i',j'|k)=\sum_{i'=1}^{D_{1}}\sum_{j'=1}^{D_{2}}p(i,j|i',j',k)p(i'|k)p(j'|k)\,,
\end{eqnarray}
and identifying $p(i'|k)=\alpha^{k}_{i'}$, $p(j'|k)=\beta^{k}_{j'}$ and $C^{k}_{ij;i'j'}=p(i,j|i',j',k)$. 

\begin{figure}[h]
    \centering
    \includegraphics[width=0.5\linewidth]{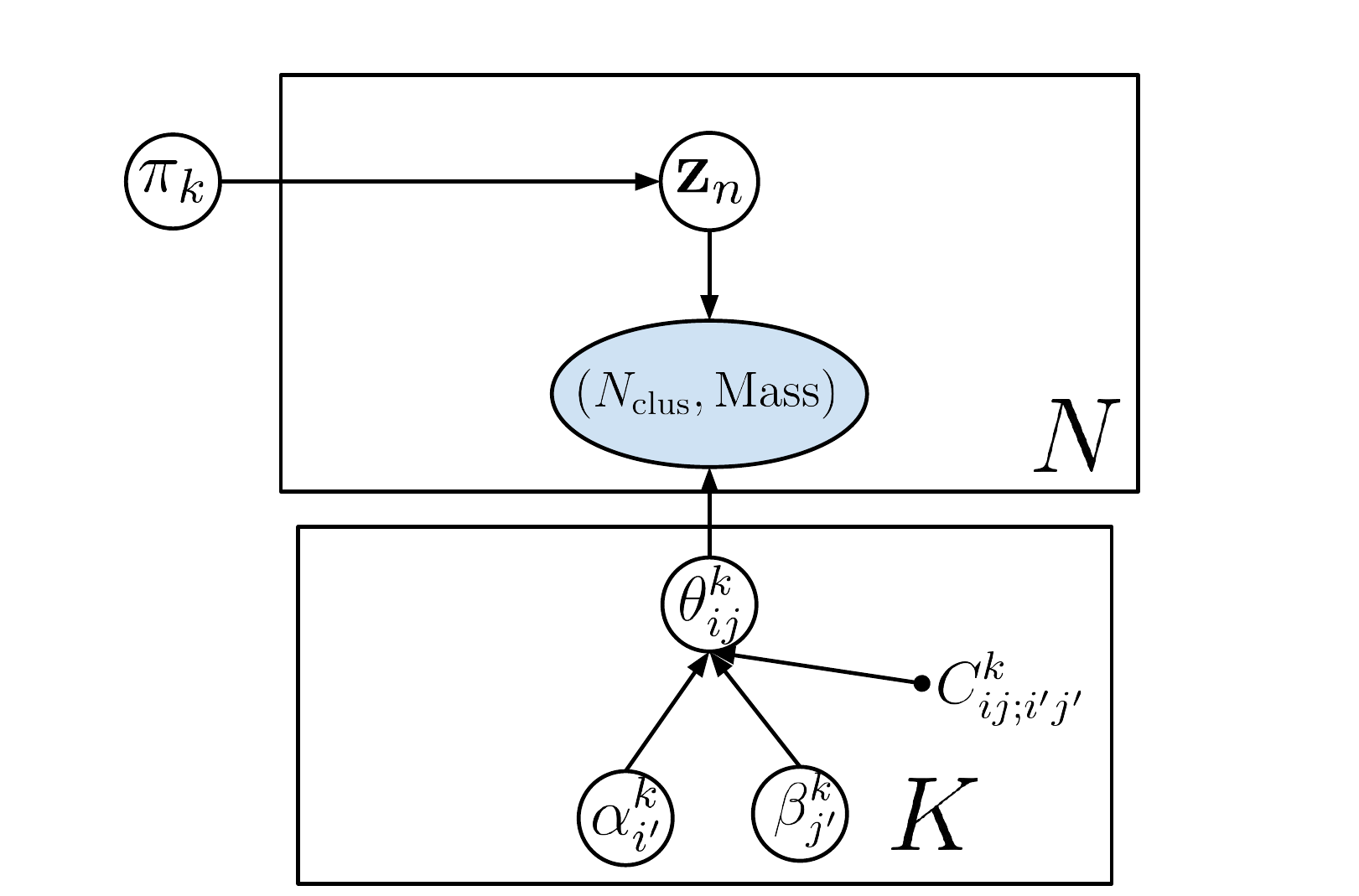}
    \caption{Graphical model for the simulation-assisted model with the addition of the transfer matrices, which are hyperparameters at this level.}
    \label{fig:graphical_model_with_correlations}
\end{figure}

The introduction of latent, unobservable, variables with conditional independence and transfer matrices that relate them to the observable values of $\vec{x}$ is at this point merely bookkeeping, with the transfer matrices acting as a generalization of covariance matrices that capture more complex relations between the two variables. Thus, if we do not impose any constraints on the transfer matrix the problem again becomes ill-posed. However, this changes if we assume in the following that the transfer matrices are fixed parameters, which we determine from Monte Carlo simulations as detailed in~\cref{sec:num_details}, and focus on inferring the posterior distribution on $\alpha^{k}_{i'}$ and $\beta^{k}_{j'}$. This assumption treats any errors in the description of the transfer matrices as subleading and allows for meaningful inference. Future efforts will be devoted to improving this approximation. The resulting graphical model is depicted in Fig.~\ref{fig:graphical_model_with_correlations} and the full likelihood can be written as
\begin{eqnarray}
    p(\boldsymbol{X}|\boldsymbol{\zeta})=\prod_{n=1}^{N}\left(\sum_{k=1}^{K}\pi_{k}\left\{\sum_{i'=1}^{D_{1}}\sum_{j'=1}^{D_{2}}C^{k}_{i_{n}j_{n};i'j'}\alpha^{k}_{i'}\beta^{k}_{j'}\right\}\right)\,,
\end{eqnarray}
which can be usefully rewritten
\begin{eqnarray}\label{eq:full_sa_likelihood}
    p(\boldsymbol{X}|\boldsymbol{\zeta})=\prod_{i=1}^{D_{1}}\prod_{j=1}^{D_{2}}\left(\sum_{k=1}^{K}\pi_{k}\left\{\sum_{i'=1}^{D_{1}}\sum_{j'=1}^{D_{2}}C^{k}_{ij;i'j'}\alpha^{k}_{i'}\beta^{k}_{j'}\right\}\right)^{N_{ij}}\,,
\end{eqnarray}
where again $N_{ij}$ is the total number of events populating bin $(i,j)$.

\subsubsection{Transfer matrices}\label{sec:num_details}

To obtain posterior samples for each considered model we utilize a Hamiltonian Monte Carlo (HMC) sampler~\cite{betancourt2018conceptualintroductionhamiltonianmonte} implemented in the \texttt{Stan}~\cite{stan} statistical software package. HMC is a particular variant of Markov Chain Monte Carlo that makes use of the derivatives of the posterior with respect to the parameters of interest to explore the relatively high-dimensional parameter space and it has been chosen due to its combination of sampling efficiency and speed. We have ensured that the sampled posterior chains pass the standard diagnostics of convergence and autocorrelation. 

For the simulation-assisted model, we estimate the transfer matrix on an additional dataset, different from data. In this ``training'' or ``Monte Carlo'' dataset, we have  to label QCD and top events and estimate the transfer matrix via an expectation-maximization (EM) algorithm that maximizes the per-class likelihood. To do so, we introduce auxiliary latent variables
\begin{eqnarray}
    \gamma_{ij;i'j'}=p(i',j'|i,j)\,,
\end{eqnarray}
that encode the probability of an event populating the bin $(i',j')$ in the sample space defined by the marginal distributions (which we denote as ``marginal bin'') given that it populates the observable bin $(i,j)$. These auxiliary variables are estimated during the E-step via the identity
\begin{eqnarray}
    \gamma_{ij;i'j'}=p(i'j'|ij)=\frac{p(i,j,i',j')}{p(i,j)}\,,
\end{eqnarray}
and are needed to estimate the marginal likelihoods $p(i'|k)=\alpha^{k}_{i'}$ and $p(j'|k)=\beta^{k}_{j'}$ and the  transfer matrix $p(i,j|i',j',k)=C^{k}_{ij;i'j'}$ during the M-step. To update the parameters of interest during the M-step, we consider the likelihood in marginal space
\begin{eqnarray}
    p(i',j'|k) = p(i'|k)p(j'|k)=\alpha^{k}_{i'}\beta^{k}_{j'}\,,
\end{eqnarray}
and assume that we have $N^{\mathrm{MC},k}_{i',j'}$ events per $(i',j')$ bin given by 
\begin{eqnarray}
    N^{\mathrm{MC},k}_{i'j'}=\sum_{i=1}^{D_{1}}\sum_{j=1}^{D_{2}}\gamma_{ij;i'j'}N^{\mathrm{MC},k}_{ij}\,,
\end{eqnarray}
where $N^{\mathrm{MC},k}_{ij}$ are the total number of events belonging to the Monte Carlo simulation of class $k$ that populate bin $(i,j)$. Thus the likelihood over the labeled dataset is
\begin{eqnarray}
    \mathcal{L}=\prod_{i'=1}^{D_{1}}\prod_{j'=1}^{D_{2}}\left(\alpha^{k}_{i'}\beta^{k}_{j'}\right)^{N^{\mathrm{MC},k}_{i'j'}}\,,
\end{eqnarray}
which can be maximized with respect to $\alpha^{k}_{i'}$ and $\beta^{k}_{j'}$ yielding the usual estimates
\begin{eqnarray}
    \alpha^{k}_{i'}&=&\frac{\sum_{j'}N^{\mathrm{MC},k}_{i'j'}}{\sum_{i'}\sum_{j'}N^{\mathrm{MC},k}_{i'j'}}=\frac{N^{\mathrm{MC},k}_{i'}}{N^{\mathrm{MC},k}}\nonumber\,,\\
    \beta^{k}_{j'}&=&\frac{\sum_{i'}N^{\mathrm{MC},k}_{i'j'}}{\sum_{i'}\sum_{j'}N^{\mathrm{MC},k}_{i'j'}}=\frac{N^{\mathrm{MC},k}_{j'}}{N^{\mathrm{MC},k}}\,,
\end{eqnarray}
where $N^{\mathrm{MC},k}$ is the total number of events for the Monte Carlo sample of class $k$. The transfer matrix is then estimated as
\begin{eqnarray}
    C^{k}_{ij;i'j'}&=&p(i,j|i',j',k)\nonumber\\
    &=&\frac{p(i',j'|i,j,k)p(i,j|k)}{p(i',j'|k)}\nonumber\\
    &=&\gamma^{k}_{ij;i'j'}\frac{N^{\mathrm{MC},k}_{ij}}{N^{\mathrm{MC},k}_{i'j'}}\,,
\end{eqnarray}
The EM algorithm is shown as a pseudo-code in~\cref{app:em_algorithm}. After training, 
%the marginal likelihoods are discarded and 
the transfer matrix can be used when inferring the conditionally independent distributions in ``marginal space'' from data. In this sense, we are treating the imperfections of the simulator as concentrated in the marginal likelihoods per-class, with the transfer matrix being correct up to subleading effects.

The code can be found at {\tt GitHub}~\cite{code-boosted-top-jets}.

\subsection{A quantifier for the goodness of the inference}\label{sec:metrics}

To assess the quality of the inference procedure we introduce the distances $D$ that quantify how far the mean values ($\overline{\pi_{k}}$, $\overline{\boldsymbol\alpha^k}$ and $\overline{\boldsymbol\beta^k}$) of the inferred parameters ($\pi_k$, $\boldsymbol\alpha^k$ and $\boldsymbol\beta^k$) are from their true values 
\begin{eqnarray}
D(\pi_{k}) & = & |\overline{\pi_{k}} - \pi_{k}^{true}| \, , \label{eq:distances1}\\
D(\alpha_i^k) & = & |\overline{\alpha_i^k} - {\alpha^{k}_{true}}_{\,i}| \, , \label{eq:distances2}\\
D(\beta_i^k) & = & |\overline{\beta_i^k} - {\beta^{k}_{true}}_{\,i}| \, . \label{eq:distances3}
\end{eqnarray}
We also compute their fluctuations such that the uncertainties of the distances $D$ are given by the standard deviations of the distributions of the inferred parameters.

Additionally, we compute the Kullback-Leibler (KL) divergence between the posterior and the data distributions both for each class and for the complete case where no labels are used. The latter case in particular allows us to get a broader picture which does not rely on knowledge of ``true'' parameters. In all cases, a smaller divergence signals a better agreement. 

Although the KL divergences provide useful metrics, they are not rigorous statistical comparisons between models. Depending on the access to the true parameters, one could use an array of more involved but powerful techniques. If the true parameters are known, posterior credible intervals could be computed and a coverage study could be performed by bootstrapping the dataset. If one does not wish to rely on access to the true parameters, a posterior predictive check could be performed, which also involves additional samplings under the learned posterior model. Thus, more rigorous tests will in general involve additional, expensive simulations. In this work, we find that the cheaper metrics provide enough information to justify the need for correlations and validate the specific modeling proposed, without the necessity of more expensive computations. However, future implementations may and probably will need to rely on these well-established suite of metrics to ensure that the learned probabilistic model is appropriate.

\subsection{Dataset}\label{sec:dataset}

For the purposes of the following sections we resort to simulations in order to benchmark our study. With the aim to facilitate a connection with previous analysis, we employ the standard Top Quark Tagging Reference Dataset \cite{kasieczka_gregor_2019_2603256}. This dataset is generated with \texttt{Pythia8}~\cite{Bierlich:2022pfr} at c.m.~energy of $14\;{\rm TeV}$ and uses the default \texttt{Delphes}~\cite{deFavereau:2013fsa} simulation of the ATLAS detector. Furthermore, it contains $1.2$M events for training, $400$K events for testing and $400$K events available for validation. The reconstruction of fat jets is performed with the anti-$k_T$ algorithm~\cite{Cacciari:2008gp} in \texttt{FastJet}~\cite{Cacciari:2011ma} setting $R=0.8$ and demanding $p_T\in(550,650)\;{\rm GeV}$ and $|\eta|<2$. For the present analysis, we select those events in the training sample which satisfy an initial selection cut in the invariant mass of the jets $m_j$ defined by $150\;{\rm GeV}<m_j<225\;{\rm GeV}$. The choice of this particular range is dictated, on the one hand, by the requirement to take into consideration all the details of the top mass peak and, on the other hand, by the necessity to restrict the amount of parameters to be optimal for the successful inference process.

\section{Results}\label{sec:results}

We next discuss the results for inferring the true top distributions $\vec{x}=\{N_{\mathrm{clus}},\mathrm{Mass}\}$ for the frameworks assuming the mixture model with and without the transfer matrices. The first goal, presented in Section~\ref{sec:correlation_matters}, is to carry out an initial comparative analysis of the two models' performance via studying the inferred posteriors as a function of the number of events in a given test sample and the choice of prior distributions. In Section \ref{sec:quality_evaluation} we assess the quality of both models with the metric defined in Section \ref{sec:metrics}. 
As detailed in section~\ref{sec:num_details}, all the posterior samples are obtained via the \texttt{Stan}~\cite{stan} statistical software package.

\subsection{Why correlations matter}\label{sec:correlation_matters}

\begin{figure}[h]
    \centering
    \includegraphics[width=1.\linewidth]{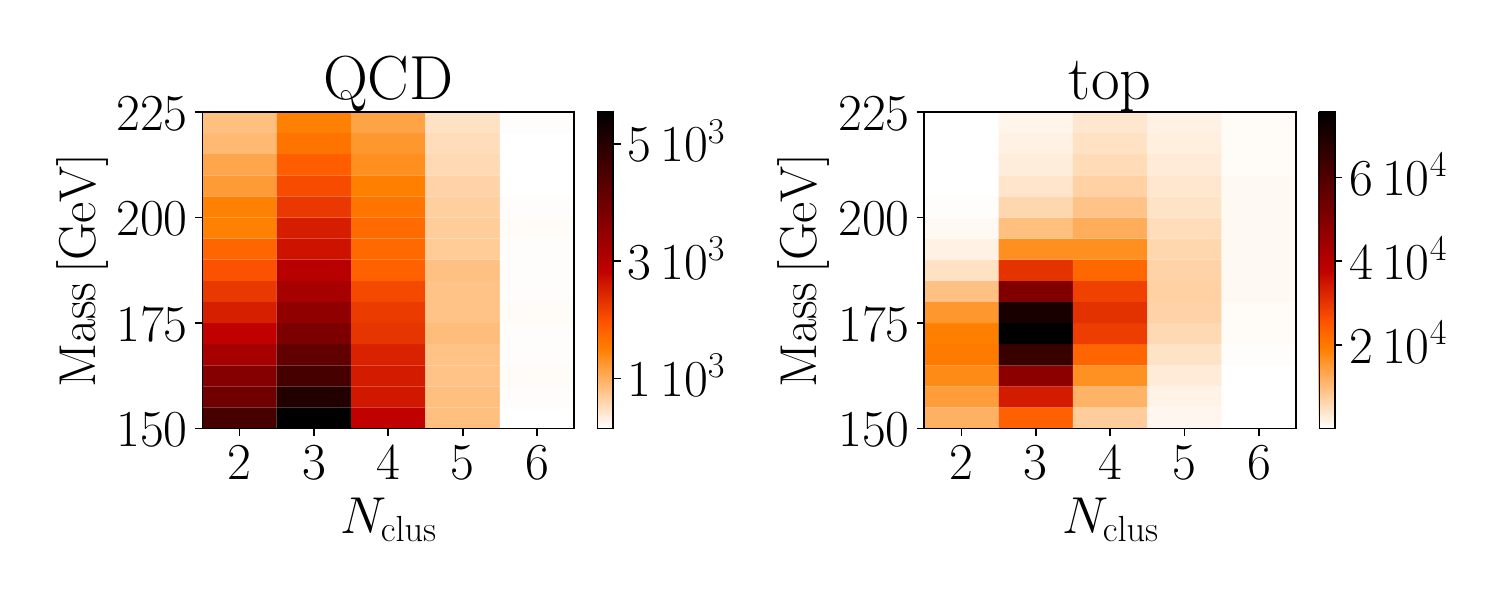}
    \caption{True two-dimensional distributions for QCD and top jets. We observe a non-negligible correlation in the top distribution.}
    \label{fig:correlation}
\end{figure}

As presented in Section \ref{sec:mixturemodel}, to include existing correlations between the mass and the number of clusters of the jet we extend the conditionally independent mixture model with transfer matrices derived from simulations. For the selection criteria and the chosen jet mass range defined in Section \ref{sec:dataset} (150-225 GeV), we plot in the right (left) panel of Fig.~\ref{fig:correlation} the top (QCD) jet mass and the number of clusters in the jet~\footnote{We have checked that discarding the bin with just one cluster in the jet is beneficial for the inference process, therefore the number of clusters will cover the range from two to six throughout the following analysis.} as 2D-histograms where the color scale indicates the number of events in a given bin, going to darker tones for larger amounts. Merely a visual inspection suggests the existence of correlations within each class. We quantify their strengths by calculating the mutual information (MI)~\cite{Bishop} and the Pearson correlation coefficient ($r$)~\cite{Lyons:1986em} presenting the results in Table~\ref{tab:mutual_info}. We see there that both variables are in fact correlated, although  the correlation is smaller for QCD jets. We also observe how MI is more sensitive than $r$, pointing towards the importance of non-linear dependencies between variables.

\begin{table}[h]
\centering
\begin{tabular}{cccccc}
\hline
MI(QCD) & MI(top) & MI(top)/MI(QCD) & $r$(QCD) & $r$(top) & $r$(top)/$r$(QCD) \\
\hline
0.008 & 0.069 &  8.6 & 0.12 & 0.36 & 3\\ 
\end{tabular}
\caption{Mutual information MI and Pearson correlation coefficient $r$ of the mass and number of clusters per jet for QCD and top jets.}
\label{tab:mutual_info}
\end{table}

As a first measure of the impact of the inclusion of correlations on the inference performance, we study the inferred probability over the mixture fraction, namely, the probability $\pi_1$ of top jets  for a given sample as a function of the number of events contained in it.  To compute the posterior, we benchmark our models with the following prior distributions for the parameters $\boldsymbol{\zeta}$: i) a uniform distribution for the probability of top jets ($\pi_1$) in the sample, ii) marginal QCD and top distributions for the number of clusters, iii) a decreasing linear approximation for the QCD jet mass distribution, and iv) a deformation of the marginal top jet mass distribution. The rationale behind this choice is to study whether a combination of agnostic and biased priors can still lead to an acceptable posterior once the algorithm sees the data. In particular, the deformation of the marginal top jet mass distribution consists in cutting off the top mass peak. This puts under stress a major discriminating feature between top and QCD jets and allows to test the robustness of the inference process.

The prior values for the fractions $\pi_k$ are generated with a uniform Dirichlet distribution
\begin{eqnarray}
{\bf\pi_{0,1}} & \sim & \mathrm{Dirichlet}(1,1) \, .
\end{eqnarray}
The prior values for the parameters $\boldsymbol\alpha^k$ and $\boldsymbol\beta^k$ are obtained by sampling other Dirichlet distributions 
\begin{eqnarray}
{\boldsymbol\alpha^k} & \sim & \mathrm{Dirichlet}({\boldsymbol\eta_{\alpha}^k}) \, , \\
{\boldsymbol\beta^k} & \sim & \mathrm{Dirichlet}({\boldsymbol\eta_{\beta}^k}) \, ,
\end{eqnarray}
where the vectors $\boldsymbol\eta_{\alpha,\beta}^k$ are given by
\begin{eqnarray}
\label{sigma}
{\boldsymbol\eta_{\alpha,\beta}^k} & = & \boldsymbol{p}_{\alpha,\beta}^k \, \Sigma \, ,  
\end{eqnarray}
with $\boldsymbol{p}_{\alpha}^k$ and $\boldsymbol{p}_{\beta}^k$ standing for the input parameters of number of clusters and mass probability distributions, respectively, and $\Sigma$ is a common normalization.     As the parameters $\boldsymbol\alpha^k$ and $\boldsymbol\beta^k$ are generated by the Dirichlet distribution, their expectation values are given by the input parameters of the multinomial distributions $\boldsymbol{p}_{\alpha,\beta}^k$
\begin{eqnarray}
\mathrm{E}[\boldsymbol\alpha^k] & = & \boldsymbol{p}_\alpha^k \, , \\
\mathrm{E}[\boldsymbol\beta^k] & = & \boldsymbol{p}_\beta^k \, ,
\end{eqnarray}
and the variance results
\begin{eqnarray}
\mathrm{Var}[\boldsymbol\alpha^k] & = & \frac{\boldsymbol{p}_\alpha^k(1-\boldsymbol{p}_\alpha^k)}{\Sigma+1} \, , \\
\mathrm{Var}[\boldsymbol\beta^k] & = & \frac{\boldsymbol{p}_\beta^k(1-\boldsymbol{p}_\beta^k)}{\Sigma+1} \, .
\end{eqnarray}
The input parameters of the multinomial distributions $\boldsymbol{p}_{\alpha,\beta}^k$ for the model with and without correlations are taken to be the parameters of the marginal distributions for the case of the number of clusters for both top and QCD, the marginal distribution with a deformation of the peak of the top jet mass distribution, and a decreasing linear approximation for the QCD mass distribution. Notice that the normalization factor $\Sigma$ controls the variation of the prior values for the parameters $\boldsymbol\alpha^k$ and $\boldsymbol\beta^k$.  Large $\Sigma$ implies ${\boldsymbol\alpha^k}$ and $\boldsymbol\beta^k$ concentrated on their mean $\boldsymbol{p}_{\alpha,\beta}^k$, and vice versa. Although the choice of $\Sigma$ is subjective, it is not arbitrary: the prior needs to encode the trust in the Monte Carlo simulations while providing an adequate range of allowed discrepancies. Because in this work we have access to the true parameter values, we can validate the choice of $\Sigma$ simply by observing that the resulting posterior can be likelihood-driven and centered around these true values for large enough number of events. When applying this method to data where the true values are unknown, a data-driven prior validation should be implemented. One possibility is to perform prior predictive checks~\cite{radev2020bayesflow} to ensure that the chosen $\Sigma$ produces realistic datasets. Another possibility is to build a hierarchical model where $\Sigma$ is promoted to a random variable and marginalized over (see Ref.~\cite{gelman2013bayesian}, Chapter 5), reducing the prior dependence at the expense of increased model uncertainty. Since the goal of this work is to validate the introduction of correlations, we leave a more detailed prior exploration for future work.

Since the transfer matrix that encodes correlations is obtained from Monte Carlo simulations, we seek to reduce the possibility of overfitting by computing the transfer matrix on a different sample from the one used for posterior inference. We also use this different sample to compute two additional sets of distributions: the distributions in marginal space $\{\alpha^{k}_{i'},\beta^{k}_{j'}\}$ that are obtained in addition to the transfer matrix, and  the marginal distributions. The latter are obtained by marginalizing over one of the variables ($\{\alpha^{k,\mathrm{marg.}}_{i}=\sum_{j=1}^{D_{2}}p(ij|k),\beta^{k,\mathrm{marg.}}_{j}=\sum_{i=1}^{D_{1}}p(ij|k)\}$) and match the distributions in marginal space only when there is no correlation (or in other words when the transfer matrix is the identity).

As detailed in previous paragraphs, the marginal distributions are used to define the prior distributions. For the model with no correlations, the marginal distributions are also the target or ``true'' distributions against which we compare both the prior and the posterior distributions. For the model with correlations, the ``true'' distributions consist of the distributions in marginal space obtained along with the transfer matrix.

Furthermore, in order to study the degree of dependence of the inferred parameters on the priors, we evaluate the inference outputs for two different allowed variations ($\Sigma=100,1400$) of the prior distributions around the corresponding mean values, which we term ``loose'' and ``tight'' accounting for the restrictions that they impose on the posterior. In addition, we aim at studying the inference performance in unbalanced samples of top and QCD jets. Therefore, we set the fraction of top jets $\pi_1=0.3$, independently of the number of events in the sample. Since we observed, as in principle expected, that a larger fraction turned out to improve the distinction between classes, we consider this relatively low top jets fraction as a conservative value to determine the goodness of the inference procedure. Much lower fractions will negatively affect the inference performance, although we have not explicitly explored the lowest sensitivity of the tagger since we expect fairly enriched samples to be available from $t\bar t$ production.

\subsubsection{Loose priors}\label{sec:loose_priors}

\begin{figure}[h]
    \centering
    \includegraphics[width=0.6\linewidth]{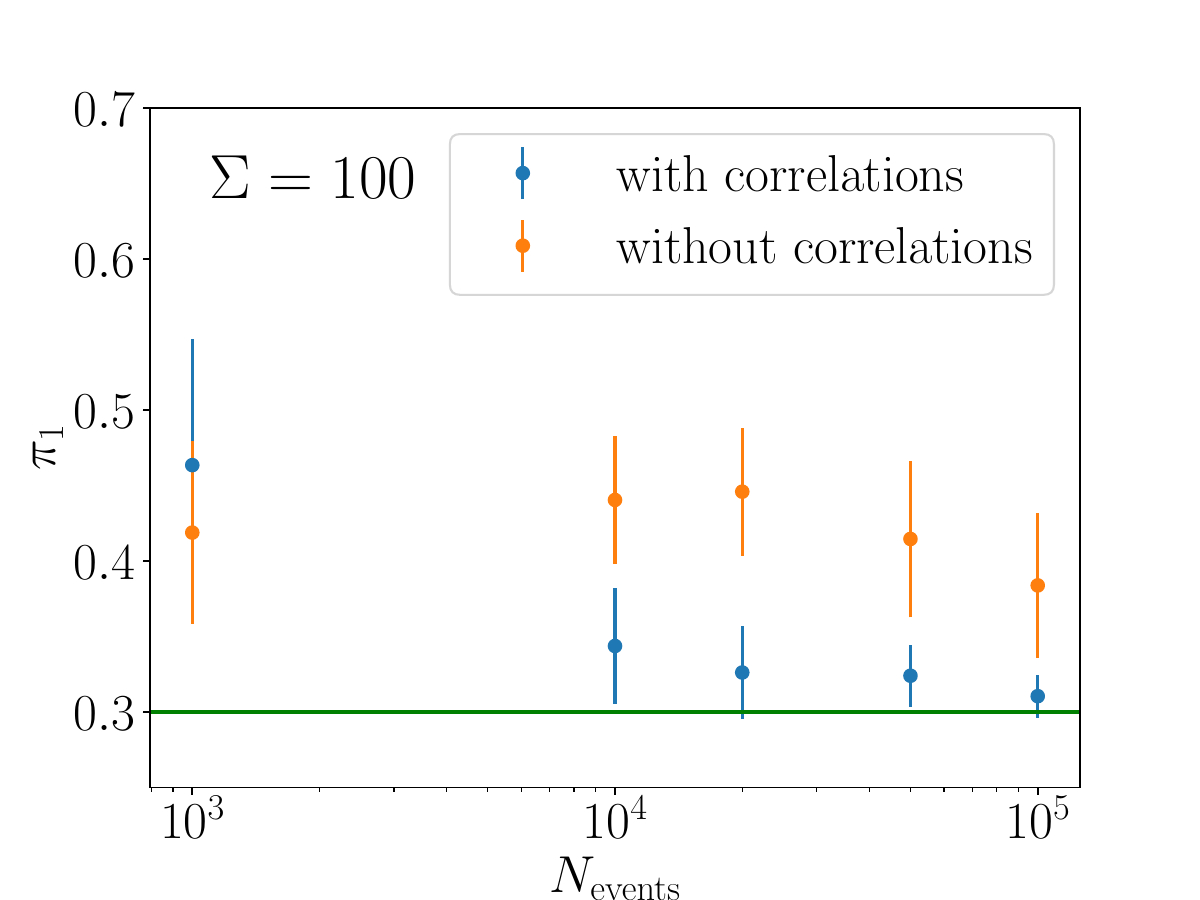}
    \caption{Inferred probability of top jets $\pi_1$ as a function of the number of events in the test samples for $\Sigma = 100 $. The blue(orange) points stand for this probability for the model with(without) correlations.}
    \label{fig:Sigma100_Nevents}
\end{figure}

We show in Fig.~\ref{fig:Sigma100_Nevents} the inferred probability of top jets as a function of the number of events in the test samples for $\Sigma=100$.
The blue (orange) points represent this probability for the mixture model with (without) correlations. The vertical error bars stand for the standard deviation over a set of 1500 samples of the posterior, and the horizontal green line corresponds to the true value of the probability of top jets ($\pi_1=0.3$).
We see that for $10^3$ events both the uncorrelated and correlated models give compatible probabilities within the uncertainties but both lie above the true value. As the number of events increases, the orange points show large fluctuations and they still keep away from $\pi_1=0.3$. Instead, the mean values of the blue points display a monotonously decreasing behaviour towards the true value. From $2\cdot10^4$ events these points are compatible within $\sim$1$\sigma$ uncertainty with $\pi_1=0.3$. Moreover, the size of the blue error bars decrease as a function of the number of events. This tendency is not verified by the orange error bars which remain roughly of the same size.   

It is worth stressing that the priors used in the inference problem do not match the true distributions in the data, and therefore the inferential process corrects the priors and approaches the trues as it sees the data.

These features suggest that, as expected, the model with correlations is more suitable to reproduce the data. As the amount of events increase, the intra-class correlations have a greater and greater impact on the inference process and the model without correlations is not capable to correctly reproduce the unlabelled data. Even so, the probability of top jets is just one parameter in the mixture models and we are equally interested in retrieving the correct distributions for the posteriors corresponding to the number of clusters and the mass of the jet.  As we show in the next paragraphs, these other distributions also exhibit a good agreement with the data for the model which includes correlations and fail in the case of the model without correlations. 

\begin{figure}[h]
\centering
\hspace*{-0.8cm}
\begin{tabular}{ccc}
\includegraphics[width=0.33\linewidth]{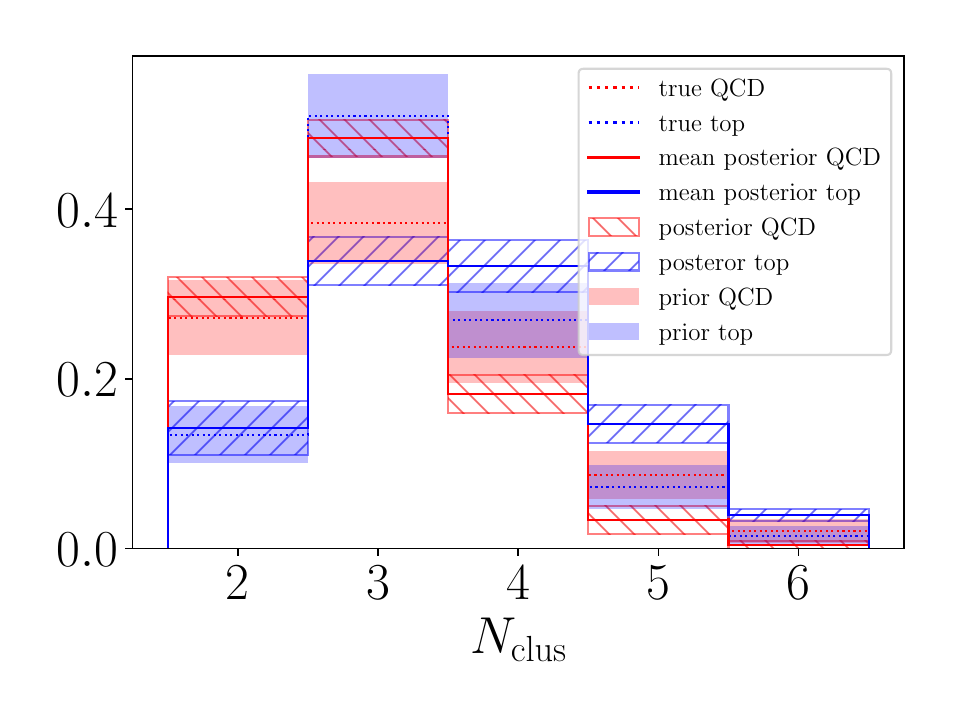} &
\includegraphics[width=0.33\linewidth]{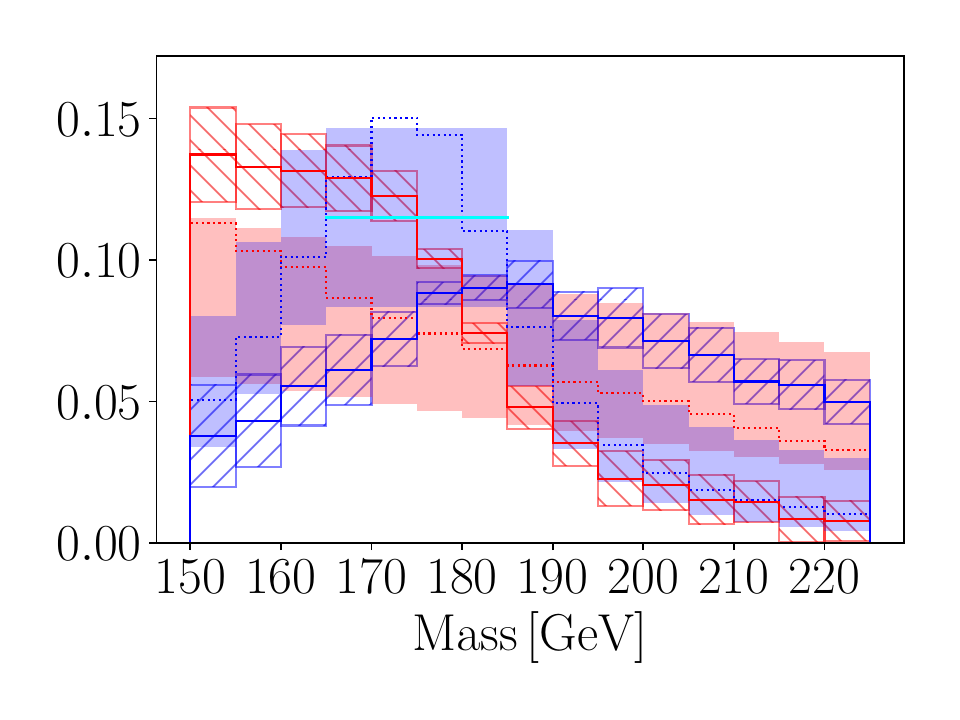} &
\includegraphics[width=0.33\linewidth]{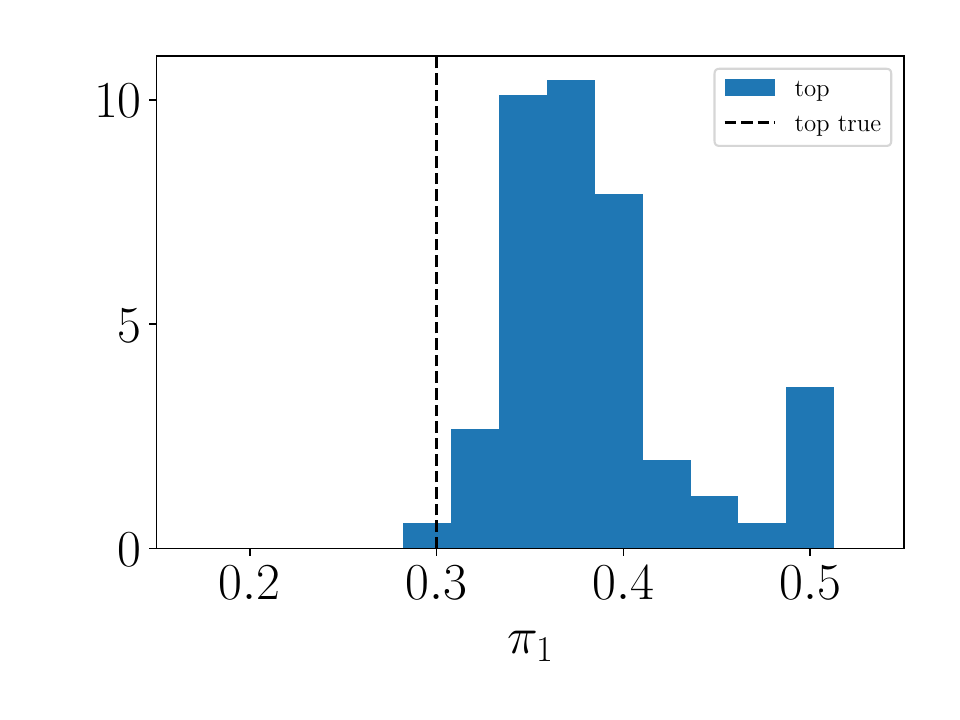} \\
\includegraphics[width=0.33\linewidth]{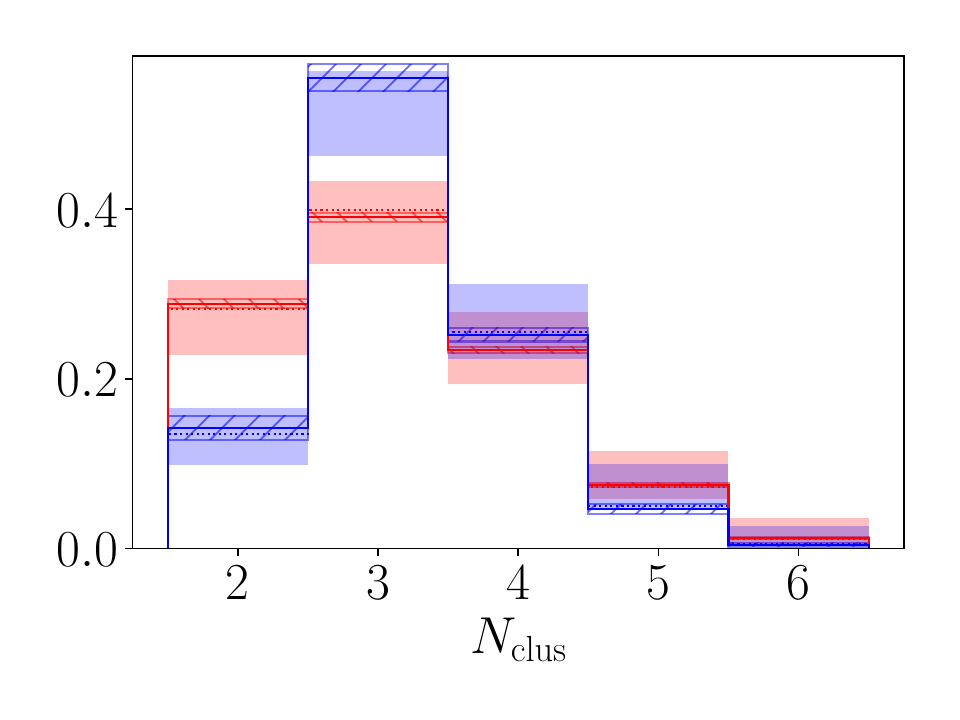} &
\includegraphics[width=0.33\linewidth]{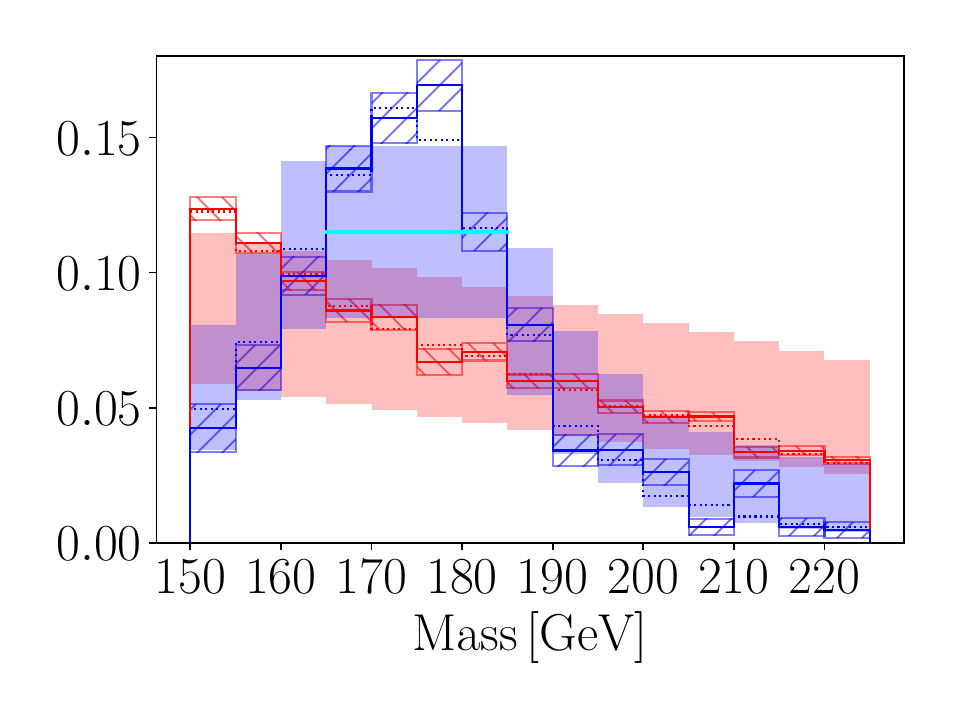} &
\includegraphics[width=0.33\linewidth]{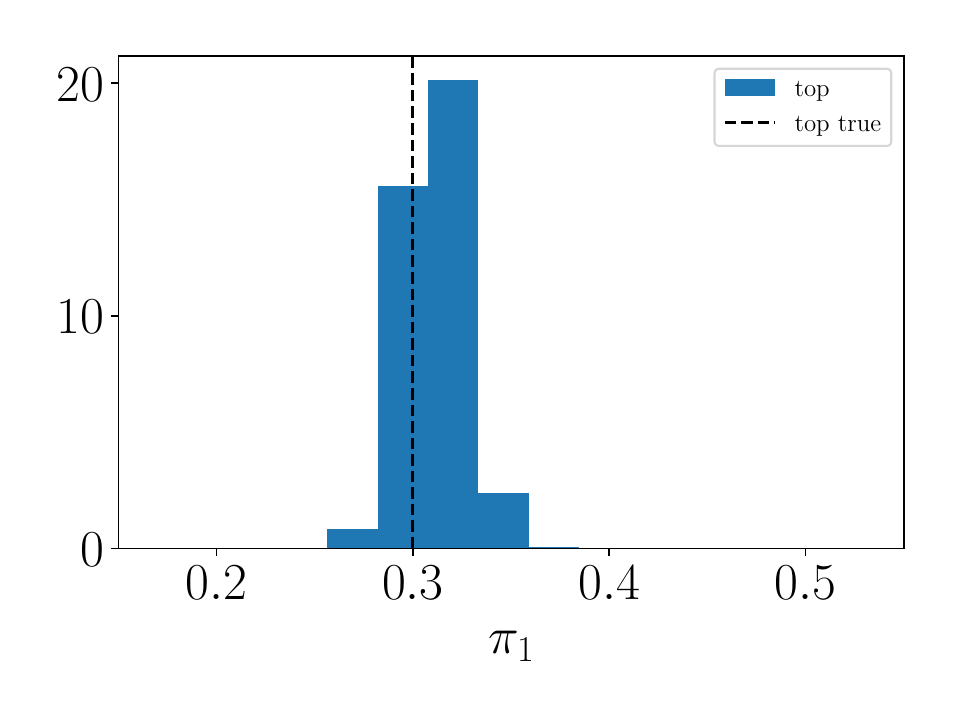} \\
\end{tabular}
\caption{Number of clusters, mass and the probability of top jets distributions for the model with (bottom row) and without (upper row) correlations with $\Sigma = 100$.  See text for the details.}
\label{fig:inference_with/without_correlation_sigma_100}
\end{figure}

For the model with(without) correlations, we then show in the bottom(upper) row, left and middle panels, of Fig.~\ref{fig:inference_with/without_correlation_sigma_100} the true values (dotted lines) together with the 1$\sigma$ prior (shaded regions) and the marginalized posterior (solid lines for the mean values and hatched regions for the standard deviations) of the number of clusters and the jet mass distributions. This figure is made for a sample of $10^5$ events in which case the posterior probability of top jets for the model with correlations reaches a closer value to the truth when compared with smaller samples. In the right panels we display the probability of top jets with the true value (vertical dashed line) and the posterior distribution in blue for both models (the prior is uniform in this case and then not shown). In agreement with Fig.~\ref{fig:Sigma100_Nevents}, we see here that the probability of top jets misses the true value for the model with no correlations, whereas it matches the truth for the model with correlations.

The plot for the number of clusters (left panel) shows that the model without correlations (upper row) is deficient in reproducing the main feature of these distributions: a higher proportion of events with three clusters for top jets associated to the decay products of the top (a $b$ quark and a $W$ with a subsequent decay into two quarks). We see that the mean values of the posteriors for top (solid blue lines) and QCD (solid red lines) jets appear inverted in that bin and there are also relatively smaller inversions in the last two bins. In contrast with this, the model with correlations reproduces well all the bins keeping particularly the correct proportion in the bin with three clusters (left panel, bottom row).

Moreover, we observe in the middle panel that the mean values of the top (solid blue lines) and QCD (solid red lines) jet mass posterior distributions closely follow the true values for the model with correlations (bottom row) but they fail in the case of the model without correlations (upper row) diminishing the top mass peak and inverting the distributions in the higher bins region. The most remarkable point here is that for the model with correlations the top mass posterior almost recovers the peak shape in spite of the deformation of the peak introduced by the prior (the mean value of the prior in the four bins surrounding the top mass is highlighted with a cyan line). Finally, the dispersion associated to the posterior of the probability of top jets, as well as the spread (hatched regions) corresponding to the posteriors of the number of clusters and the jet mass, are smaller for the model with correlations than for the model without correlations.  This is to be expected, since the model without correlations does not correspond to the data and therefore can find many unphysical combinations of parameters that explain the data fairly well without a huge drop-off in the likelihood, {\it i.e.} there is no privileged true parameter choice from which any parameter variations result in a tremendous likelihood drop. This enhancement in good parameter samples increases the uncertainties over parameter space.

The same abundance of unphysical parameter choices for the model without correlations causes the inference process to find two disconnected regions in parameter space which are related by an inversion in the class labels. That is, some posterior samples show a ``label-switching'' where the two inferred classes are inverted, disregarding the prior distribution. To account for this unphysical phenomenon, the reported posterior samples are the result of a post-processing with a criterion based on prior knowledge of the problem:
every time we get $\pi_1>0.5$, we redefine $\pi_1 \to 1 - \pi_1$ and re-label the learned per-class distributions, with the final $\pi_{1}\in[0,0.5]$. This kind of workaround to avoid label-switching is a common practice for Mixture Models~\cite{10.1111/1467-9868.00265}.

In conclusion, when the mixture model includes correlations the posteriors not only better approach the truth but also end up being more precise due to a lower variance of the posterior~\footnote{It may be interesting to notice that this conclusion is along the lines of what occurs in Ref.~\cite{cmscollaboration2025developmentsystematicuncertaintyawareneural}, where the goal of reducing the variance is linked to the best modeling of the systematics in the data.}.

\subsubsection{Tight priors}\label{sec:tight_priors}

\begin{figure}[h]
    \centering
    \includegraphics[width=0.6\linewidth]{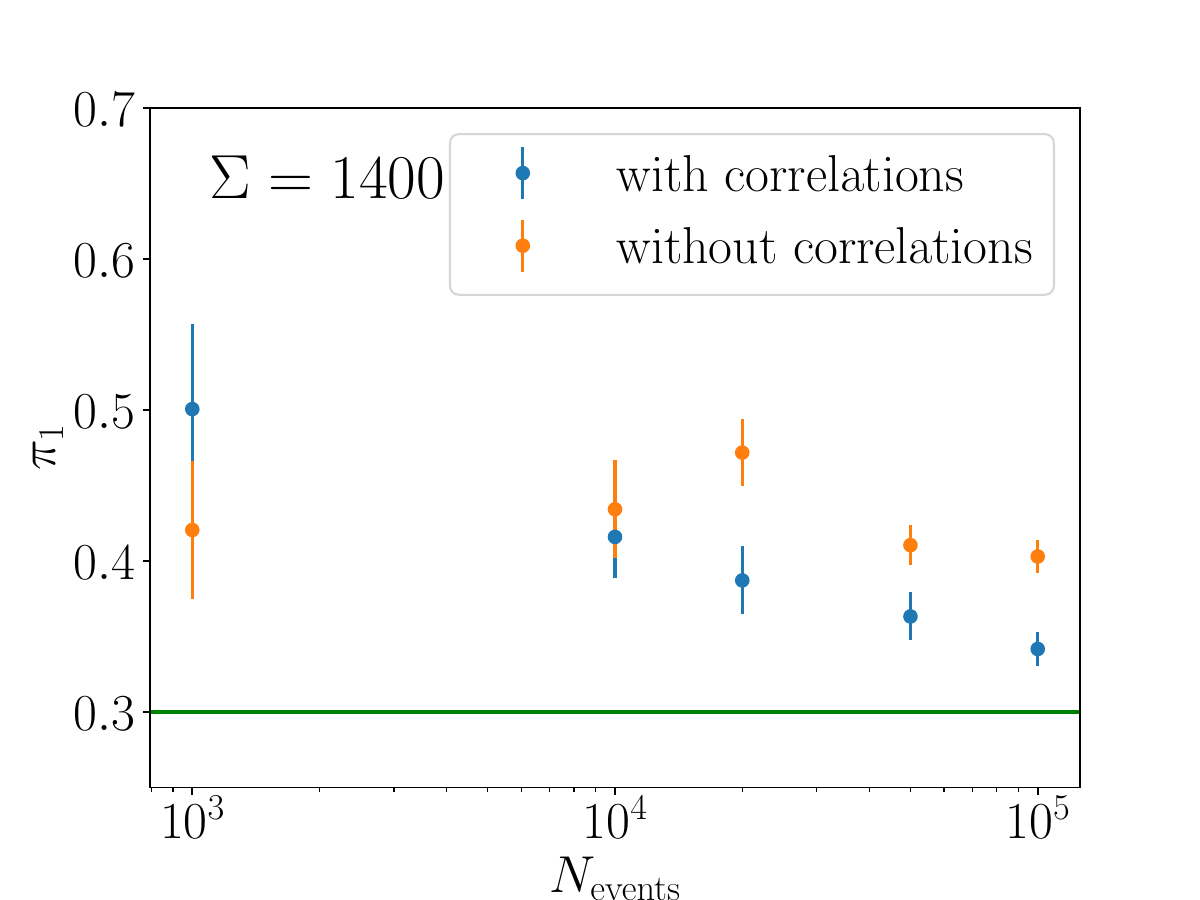}
    \caption{The inferred probability of top jets $\pi_1$ as a function of the number of events in the test samples for $\Sigma = 1400 $. The blue(orange) points stand for this probability for the model with(without) correlations.}
    \label{fig:Sigma1400_Nevents}
\end{figure}

We analyze now the effects of considering more restricted priors in the inference process. We study the case in which the multiplicative factor in the Dirichlet arguments in Eq.~\eqref{sigma} is large, namely $\Sigma=1400$.  This large factor yields that the draws from the Dirichlet are more concentrated around its mean.   In Fig.~\ref{fig:Sigma1400_Nevents} we show the inferred probability of top jets as a function of the number of events in the test samples. As in Fig.~\ref{fig:Sigma100_Nevents} the blue(orange) points depict the probability of top jets for the mixture model with(without) correlations. The vertical error bars denote the standard deviation over a set of 1500 values of the posterior, and the horizontal green line corresponds to the true value of the probability of top jets ($\pi_1=0.3$).

As in the case of $\Sigma=100$, we observe that for $10^3$ events the uncorrelated and correlated models give compatible probabilities within the uncertainties but both lie above the true value. In contrast with the loose prior, however, the mean value for the model without correlations is closer to the true value. In spite of this, from $10^4$ events on, the mean values of the blue points show again a decreasing behaviour towards the true value whereas the orange points exhibit a less pronounced slope and slightly larger fluctuations. Even considering this decreasing behaviour, the blue points are not compatible within the uncertainty with $\pi_1=0.3$, contrary to what we found for $\Sigma=100$, and this still occurs for $10^5$ events. The uncertainties also diminish as the number of events increases and the model with correlations ends up being relatively precise but not too accurate. Given that we expect more dependence on the priors compared to the loose prior case even for a relatively large number of events, it is not altogether surprising that the model with correlations is close but misses the truth because the transfer matrix is approximate and the allowed variation of the priors is more restrictive than in the case of $\Sigma=100$. Despite that, we will see in the following that the posterior distributions corresponding to the number of clusters and the mass of the jet agree with data visibly better for the model which includes correlations than for model without correlations. \\

\begin{figure}[h]
\centering
\hspace*{-0.8cm}
\begin{tabular}{ccc}
\includegraphics[width=0.33\linewidth]{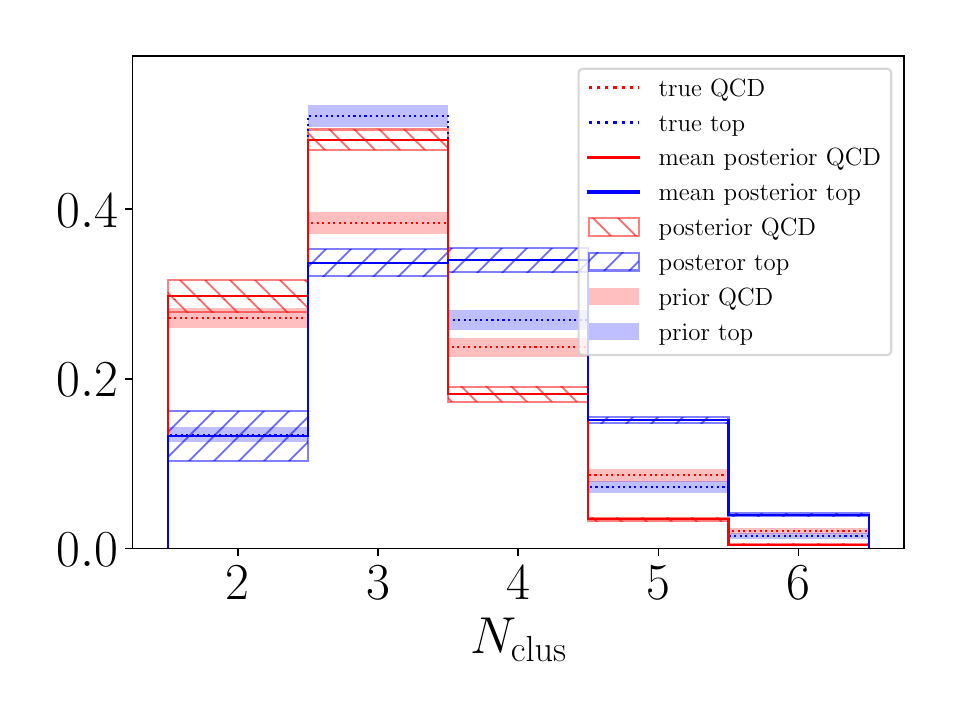} &
\includegraphics[width=0.33\linewidth]{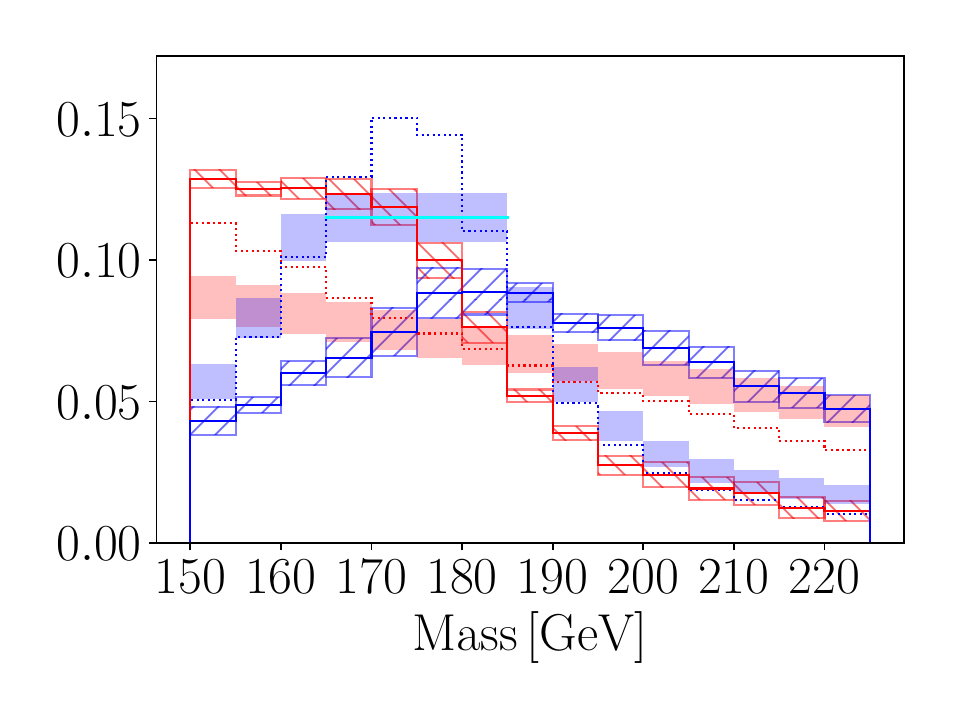} &
\includegraphics[width=0.33\linewidth]{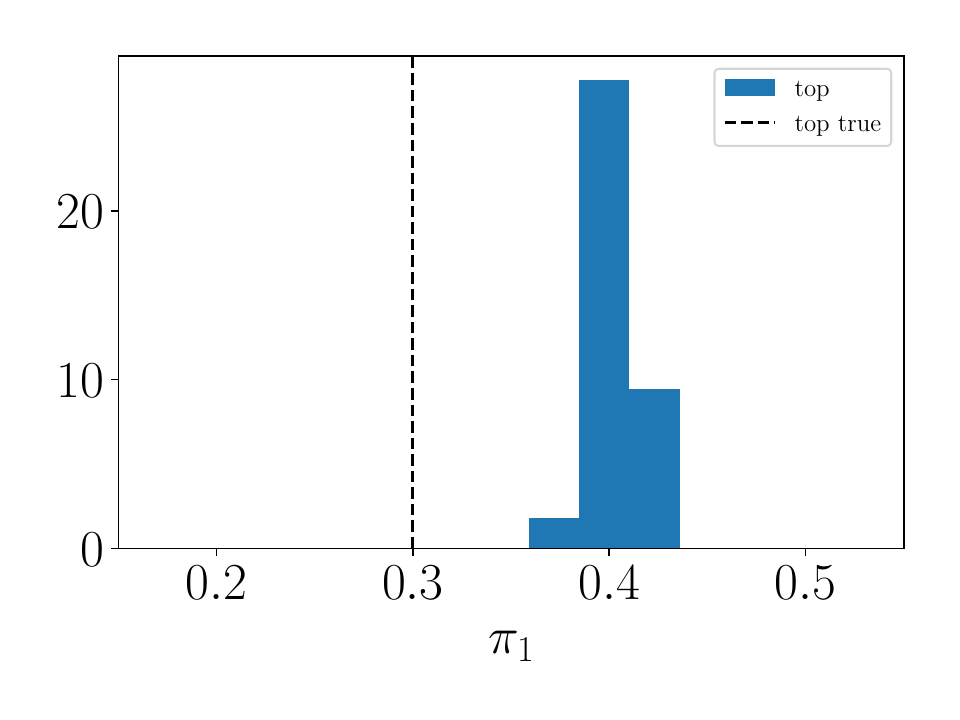} \\
\includegraphics[width=0.33\linewidth]{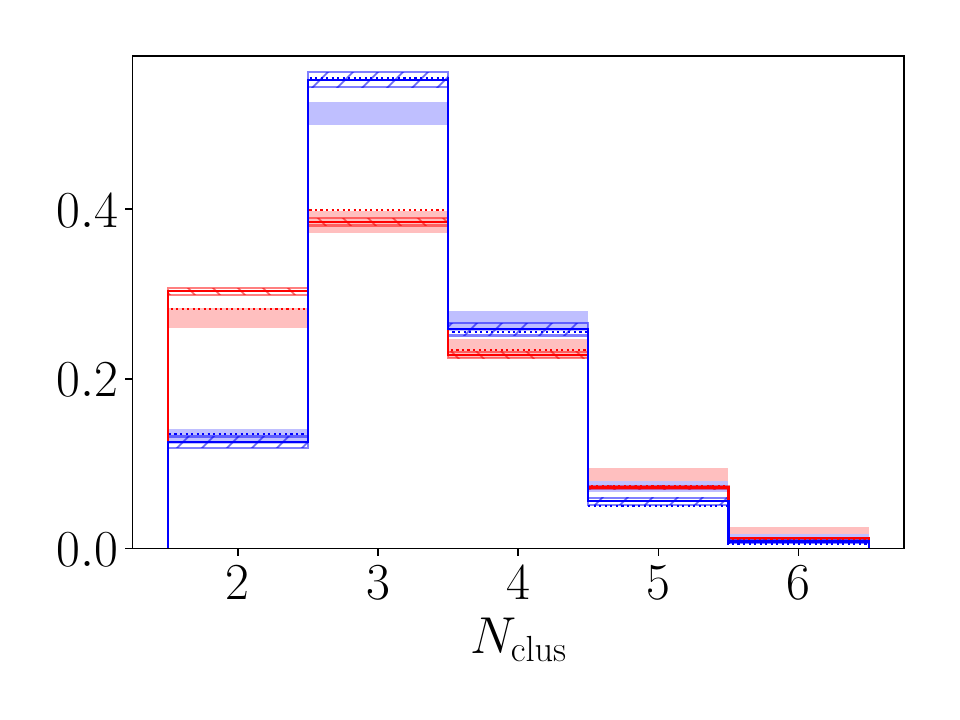} &
\includegraphics[width=0.33\linewidth]{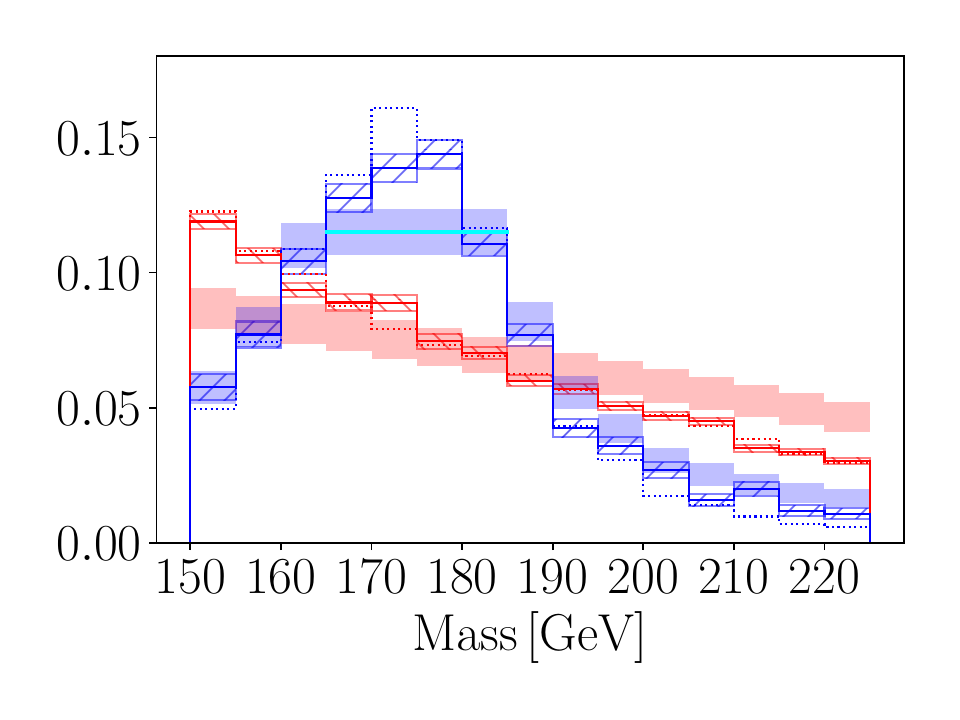} &
\includegraphics[width=0.33\linewidth]{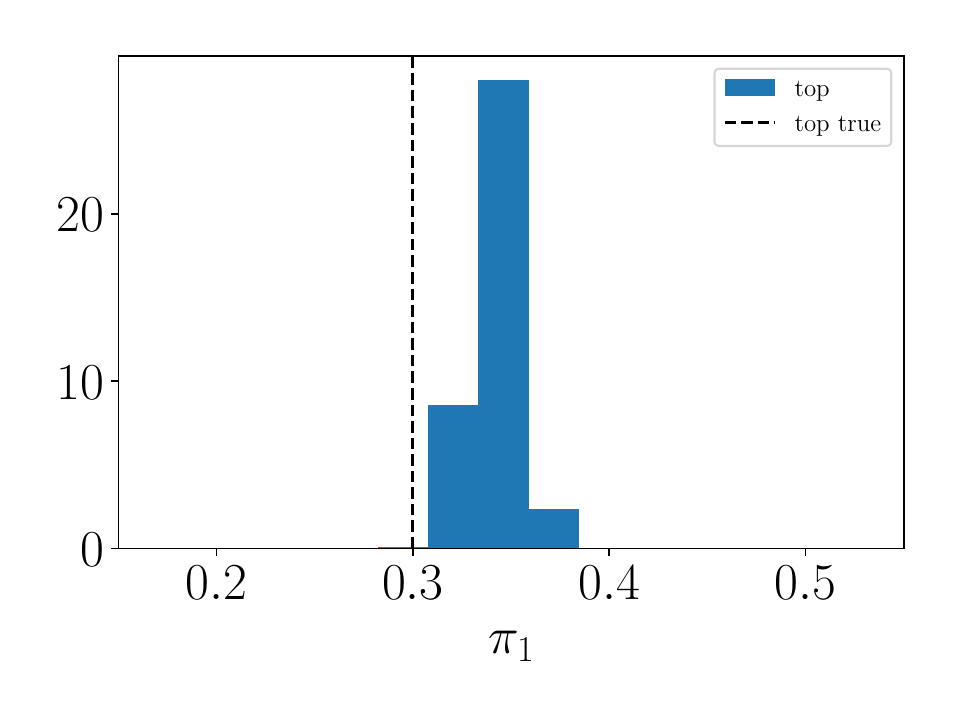} \\
\end{tabular}
\caption{Number of clusters, mass and the probability of top jets distributions for the model with (bottom row) and without (upper row) correlations with $\Sigma = 1400$.  See text for the details.}
\label{fig:inference_with/without_correlation_sigma_1400}
\end{figure}

Fig.~\ref{fig:inference_with/without_correlation_sigma_1400} corresponds to $\Sigma=1400$ and it is similar to Fig.~\ref{fig:inference_with/without_correlation_sigma_100}. For the model with(without) correlations, in the bottom(upper) row, left and middle panels, we show the true values (dotted lines) together with the prior (shaded regions) and the marginalized posterior (solid lines for the mean values and hatched regions for the standard deviations) of the number of clusters and the jet mass distributions. Likewise with loose priors, Section \ref{sec:loose_priors}, this figure is built with a sample of $10^5$ events in which case the value of the posterior probability of top jets for the model with correlations is closer to the truth when compared to smaller samples. In the right panel we display the probability of top jets with the true value (vertical dashed line) and the posterior distribution in blue for both models (the prior is again uniform in this case and not shown in the figure). In accordance with Fig.~\ref{fig:Sigma1400_Nevents}, we observe that the probability of top jets misses the true value for the model without correlations, whereas it approaches the truth for the model with correlations. 

As was the case for the loose priors, for the number of clusters (left panel) the model without correlations (upper row) fails to recreate the bin with three clusters in the jet. The posteriors for top (solid blue lines) and QCD (solid red lines) jets appear inverted there and with smaller inversions in the last two bins. The model with correlations does reproduce closely all the bins (left panel, bottom row). Additionally, the behavior of the mean values of the top (solid blue lines) and QCD (solid red lines) jet mass posterior distributions is similar to the $\Sigma=100$ case. They nearly approximate the true values for the model with correlations (bottom row) but they are unsuccessful in the case the model without correlations (upper row) where the top mass peak is reduced and the distributions in the higher bins region are inverted. In contrast to $\Sigma=100$, however, the dispersion associated to the posterior of the probability of top jets, as well as the spread (hatched regions) corresponding to the posteriors of the number of clusters and the jet mass, is comparable for the models with and without correlations. The tighter prior is regularizing the inference so that many of the unphysical parameter choices become more disfavored and thus do not contribute as significantly to the uncertainty. For the same reason, we observe an absence of switching between the probability of top and QCD jets for $\Sigma=1400$. Here again, for the model with correlations, the remarkable signature is that the top mass posterior manages to approach the peak shape even if the model is much more constrained by the large deformation of the peak (the mean value of the prior in the four bins surrounding the top mass is highlighted again with a cyan line). 

In summary, the use of tight priors has shown, as expected, that the model without correlations keeps failing the true distributions. More importantly, for the model with correlations, it has been helpful to test the robustness of the inference process when biased priors are kept away from the truth.   

\subsubsection{Prior comparison}

\begin{figure}[h]
    \centering
    \includegraphics[width=0.6\linewidth]{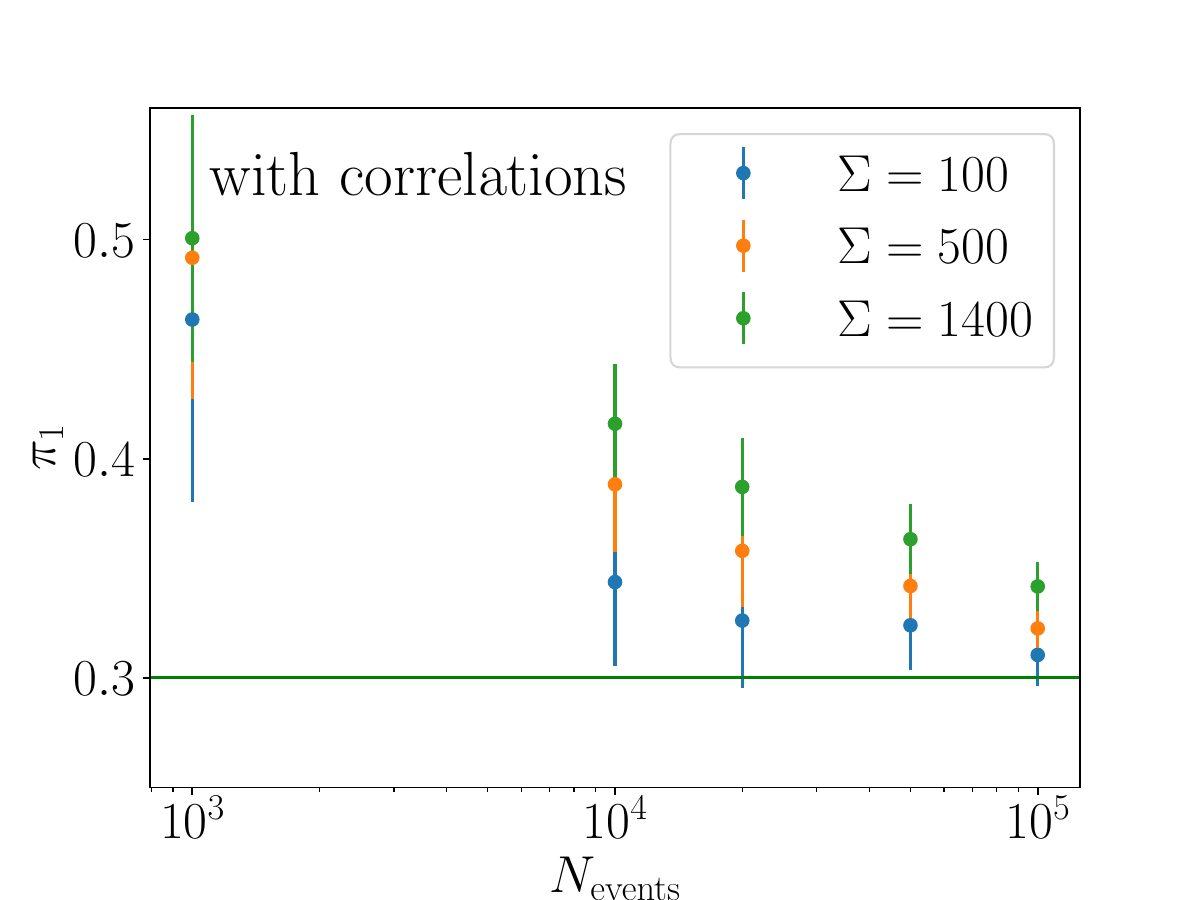}
    \caption{The inferred probability of top jets $\pi_1$ in the model with correlations as a function of the number of events in the test samples for different allowed variations of the priors.}
    \label{fig:comparison_priors}
\end{figure}

As a direct comparison between the loose ($\Sigma=100$) and tight ($\Sigma=1400$) priors for the model with correlations, we show in Fig.~\ref{fig:comparison_priors} the inferred probability of top jets as a function of events in a given sample. We also add a third possibility considering an intermediate case with $\Sigma=500$ to browse the stability of different allowed variations of the priors. For all cases, we observe how if the number of events is too small the prior dominates and biases the inferred probability. The number of events needed to overcome the prior bias depends on the choice of $\Sigma$.

In the previous sections we came to the conclusion that the model with correlations adequately captures the data distributions for both the loose and tight priors. However, one should note that the tight prior still biases all distributions even for $10^5$ events while the loose prior systematically allows the posterior probabilities to better fit the data once the amount of events is enough to dominate the inference process. The prior with $\Sigma=500$ displays a smooth progression between the other two values. 

\subsection{Model quality evaluation}\label{sec:quality_evaluation}

This section aims at evaluating the quality of both uncorrelated and correlated models through the metric defined in Eqs.~\ref{eq:distances1}-\ref{eq:distances3}. We compute the distance ($D$) between the values of the inferred parameters (the probability of top jets in the sample, the per-class probability of a certain number of clusters in the jet and the per-class probability of a given bin of jet mass) and their true values. For the model without correlations, we consider a single prior normalization, while for the model with correlations we study the impact of changing the normalization and consider both loose and tight priors as defined in section~\ref{sec:correlation_matters}. In all the cases we quantify the performance of the models employing samples of $10^5$ events.

\begin{figure}[h]
\begin{tabular}{ccc}
\includegraphics[width=6cm]{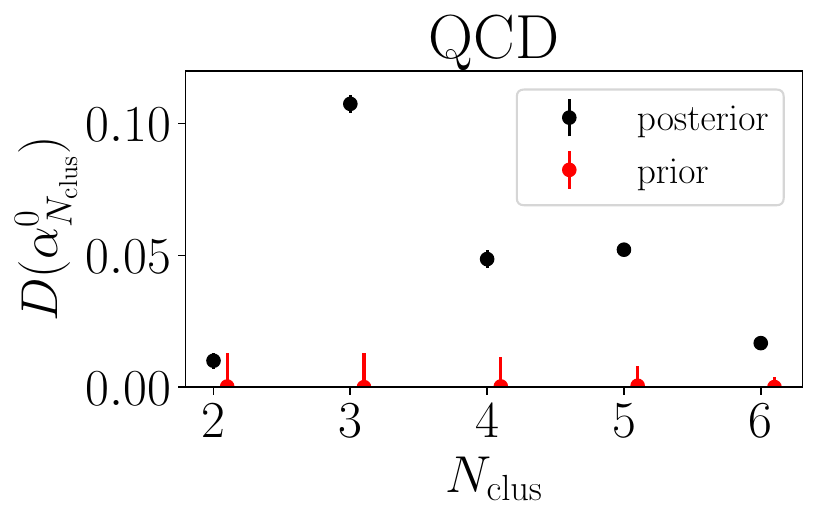}
&
\includegraphics[width=1.7cm]{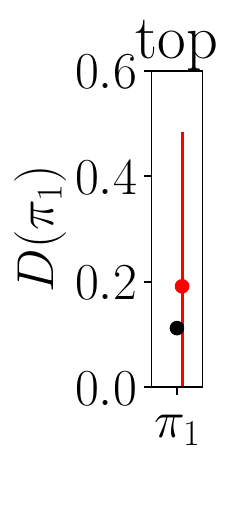}
&
\includegraphics[width=6cm]{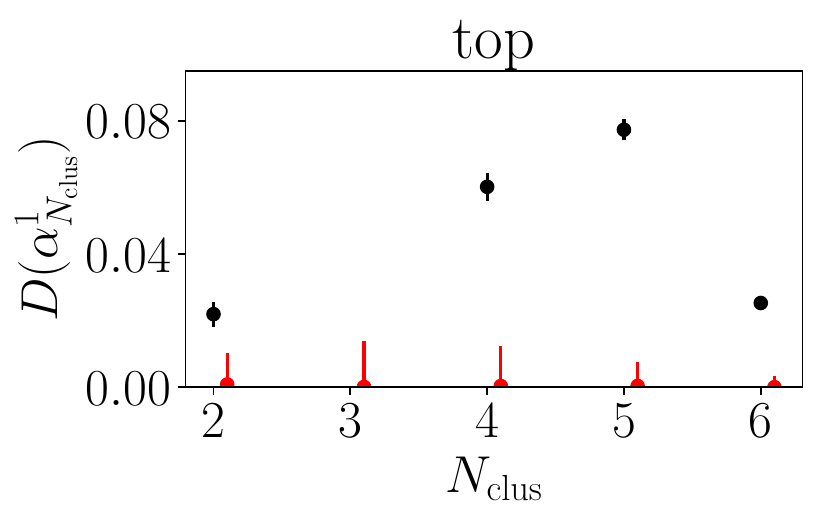}\\
\includegraphics[width=6cm]{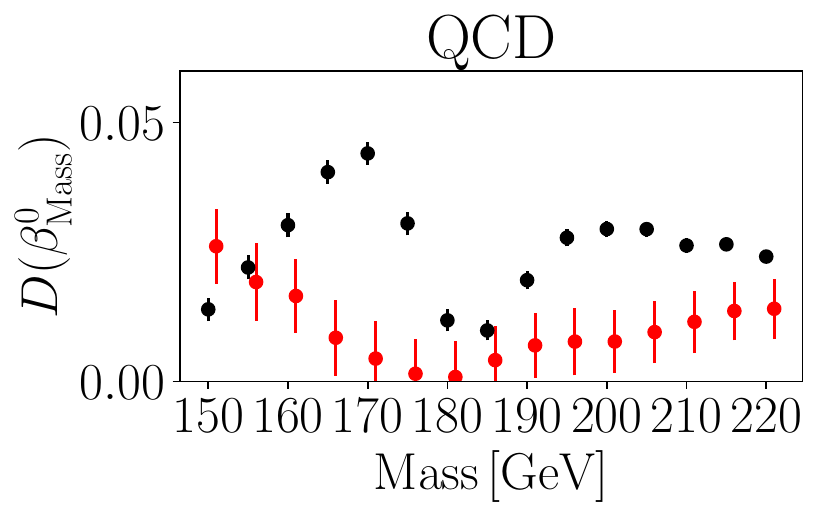}
&
&
\includegraphics[width=6cm]{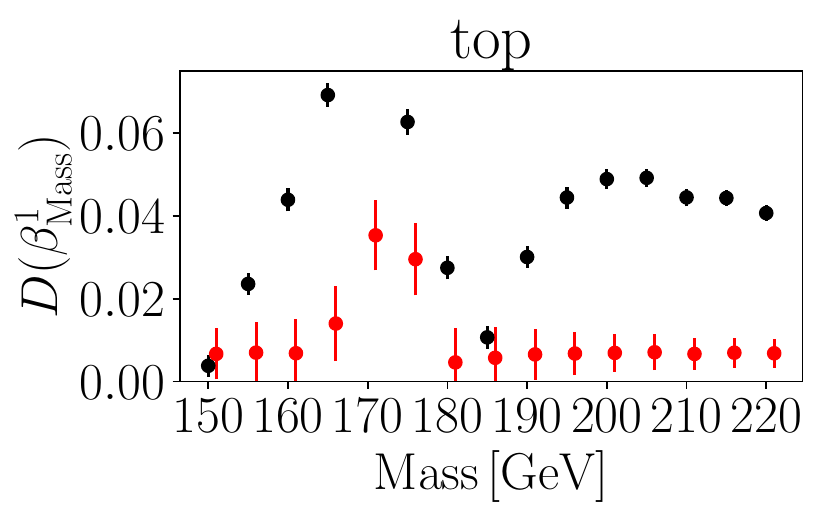}\\
&
\,
\end{tabular}
\caption{Distances of the mean values of the inferred (black circles) and prior (red circles) parameters $\pi_k$, $\boldsymbol{\alpha}^k$ and $\boldsymbol{\beta}^k$ from their true values for the model without correlation and $\Sigma = 1400$.}
\label{fig:distances_without_correlation}
\end{figure}

We show the performance of the model without correlations and tight priors ($\Sigma=1400$) in Fig.~\ref{fig:distances_without_correlation}, where the black points stand for the distance between the mean of the posterior parameters and their true values. The left(right) top panel depicts the distance for the QCD(top) number of clusters. Along with the number of clusters of top jets, we display the distance for the probability of top jets in the sample. The left(right) bottom panel contains the distance corresponding to the QCD(top) mass distribution. Additionally, the red points represent the distances between the mean of the prior parameters and their true values. Even considering standard deviations, this figure shows for all the variables that the distances for the posterior parameters are consistently larger than the corresponding ones for the prior parameters~\footnote{This is not true for all bins, with a couple of bins in the mass distributions, nor for the probability of top jets showing better agreement at the posterior level. In the latter case, the spread in the prior values of this probability is relatively large due to the fact that they are generated with a uniform distribution.}. This is in complete agreement with the results of the previous section where we conclude that the lack of correlations in the model prevents the inference from obtaining posterior distributions compatible with the truth. This conclusion remains unchanged when $\Sigma=100$ is employed in the inference, the only difference being an expected larger dispersion of the distances for the prior parameters and also a significant fluctuation in the distances of the posterior parameters due to the increased abundance of unphysical solutions for the uncorrelated model.

\begin{figure}[h]
\begin{tabular}{ccc}
\includegraphics[width=6cm]{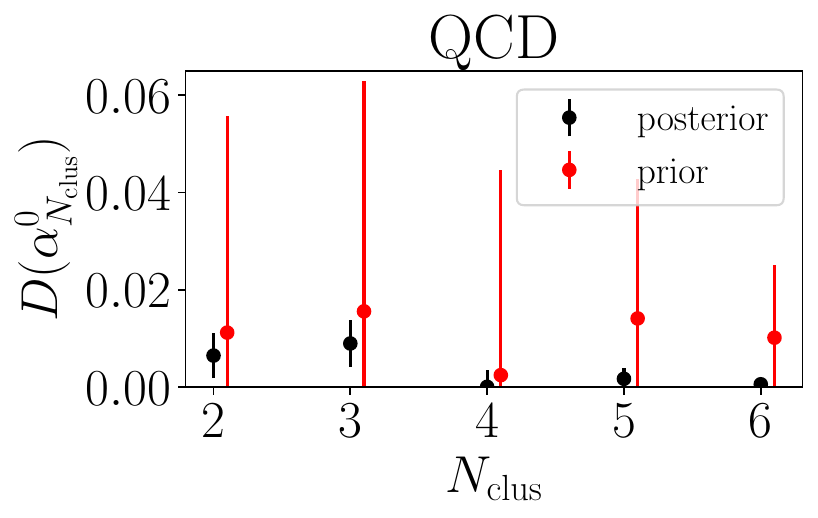}
&
\includegraphics[width=1.7cm]{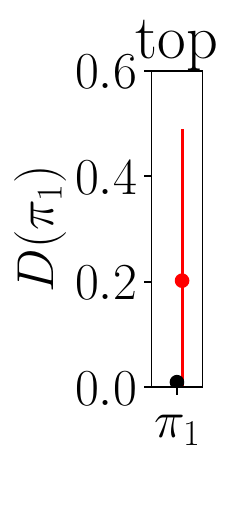}
&
\includegraphics[width=6cm]{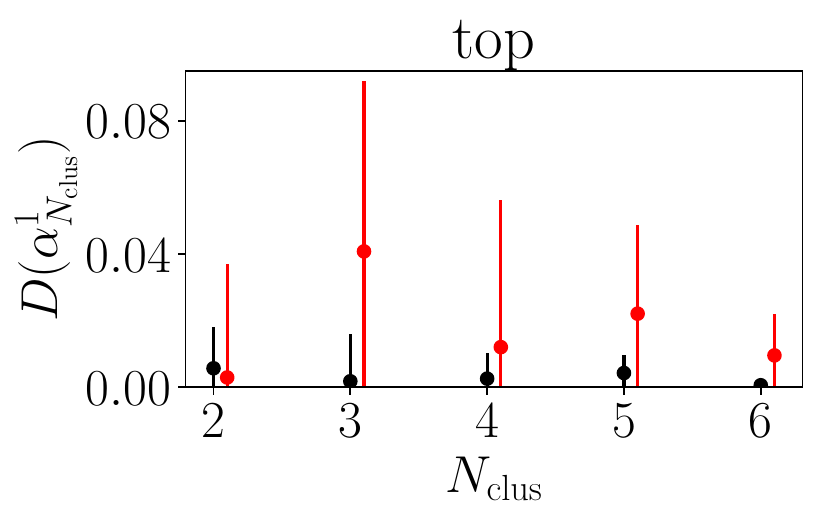}\\
\includegraphics[width=6cm]{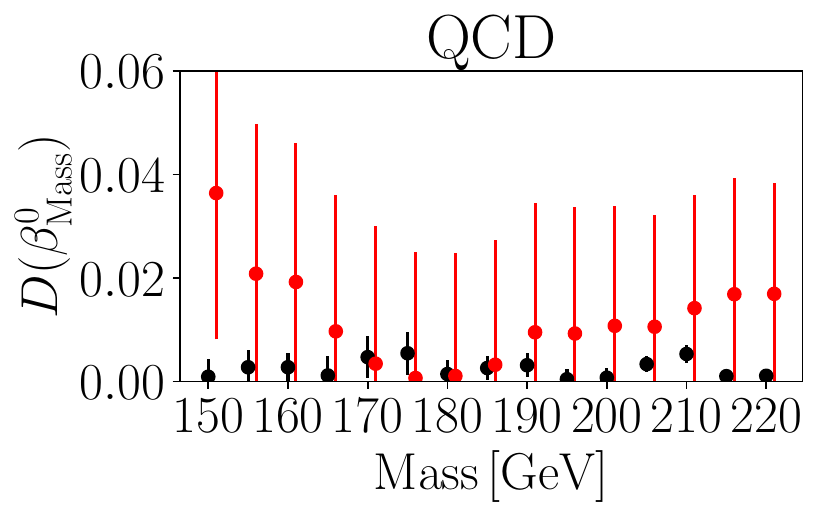}
&
&
\includegraphics[width=6cm]{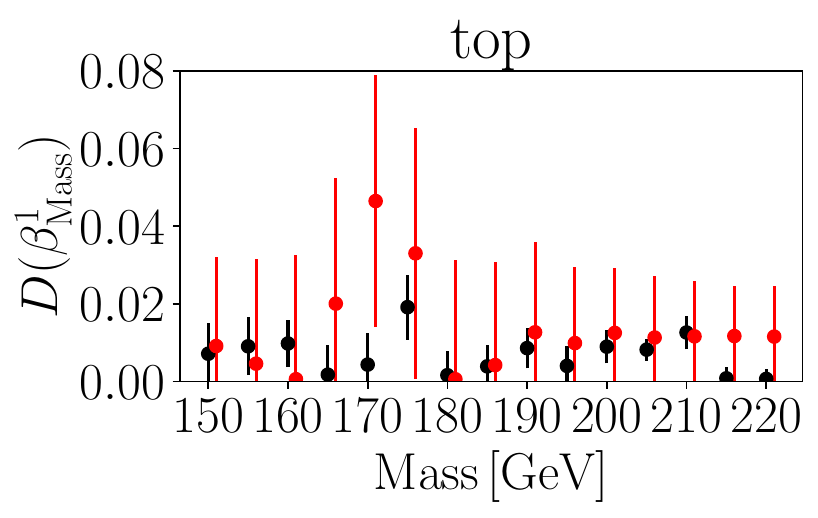}\\
&
\,
\end{tabular}
\caption{Distances of the mean values of the inferred (black circles) and prior (red circles) parameters $\pi_k$, $\boldsymbol{\alpha}^k$ and $\boldsymbol{\beta}^k$ from their true values for the model with correlation and $\Sigma = 100$.}
\label{fig:distances_with_correlation_Sigma_100}
\end{figure}

\begin{figure}[h]
\begin{tabular}{ccc}
\includegraphics[width=6cm]{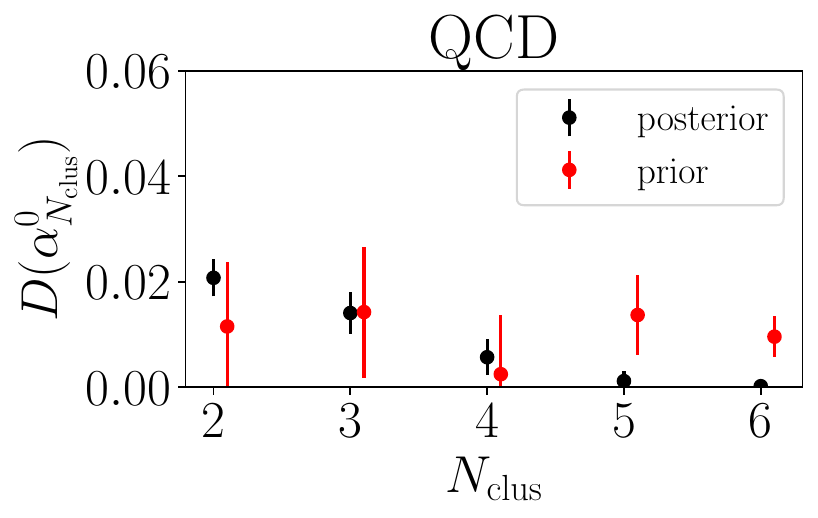}
&
\includegraphics[width=1.7cm]{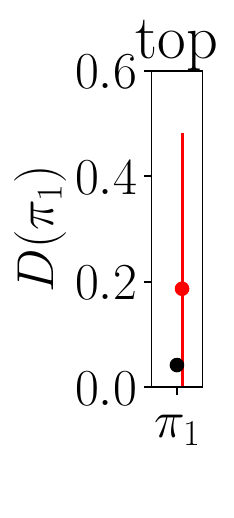}
&
\includegraphics[width=6cm]{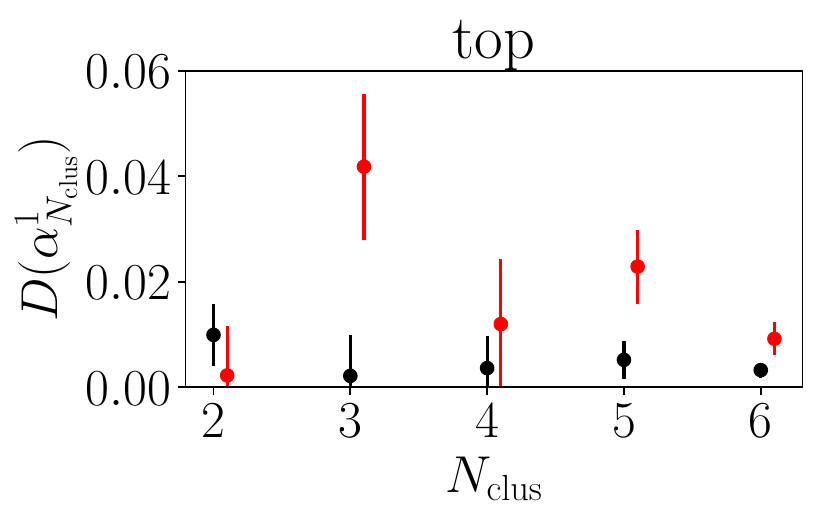}\\
\includegraphics[width=6cm]{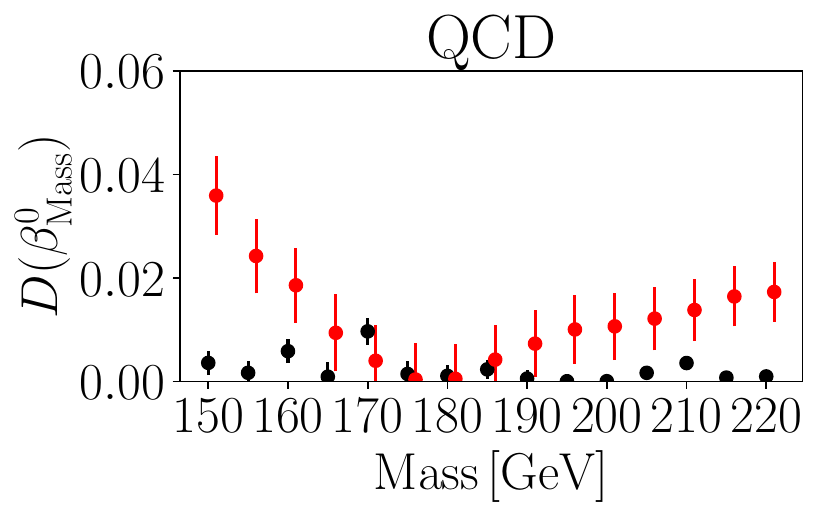}
&
&
\includegraphics[width=6cm]{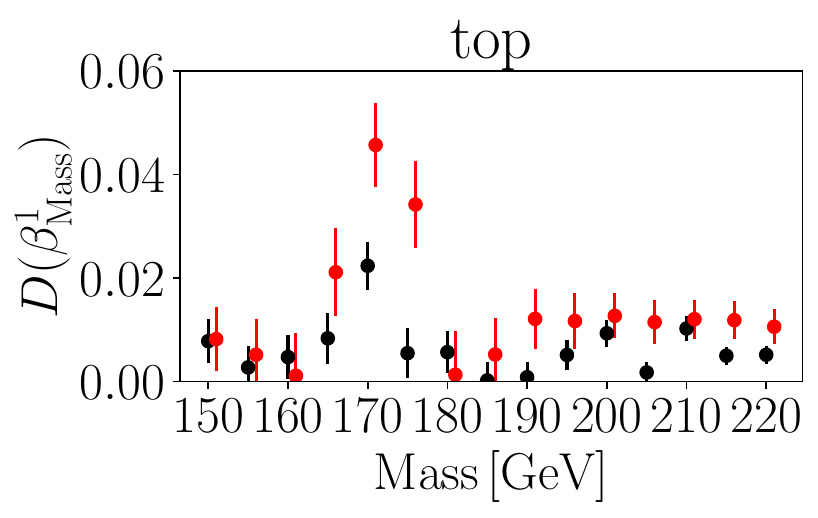}\\
&
\,
\end{tabular}
\caption{Distances of the mean values of the inferred (black circles) and prior (red circles) parameters $\pi_k$, $\boldsymbol{\alpha}^k$ and $\boldsymbol{\beta}^k$ from their true values for the model with correlation and $\Sigma = 1400$.}
\label{fig:distances_with_correlation_Sigma_1400}
\end{figure}

Figs.~\ref{fig:distances_with_correlation_Sigma_100} and \ref{fig:distances_with_correlation_Sigma_1400} correspond to the model with correlations for loose ($\Sigma=100$) and tight ($\Sigma=1400$) priors, respectively. Like Fig.~\ref{fig:distances_without_correlation}, they depict in black(red) points the distance between the mean of the posterior(prior) parameters and their true values. As a general pattern, we see in Fig.~\ref{fig:distances_with_correlation_Sigma_100} that the distances for the mean values of the posterior parameters are smaller than the corresponding ones for the prior parameters with the exception of some few bins in the cluster and mass distributions. Additionally, the black points have in all cases much smaller fluctuations than the red points. The metric quantifies in this way not only the improvement of the posteriors with respect to the priors but also the performance observed in Section \ref{sec:loose_priors} for the model with correlations where we find that the inferred parameters approach the truth with relatively small uncertainties.  Additionally, by comparing Figs.~\ref{fig:distances_with_correlation_Sigma_100} and \ref{fig:distances_with_correlation_Sigma_1400} with Fig.~\ref{fig:distances_without_correlation} we also confirm the better performance of the model with correlations~\footnote{Notice that the uncertainties of the red points seem to be smaller in Fig.~\ref{fig:distances_without_correlation} than in Fig.~\ref{fig:distances_with_correlation_Sigma_1400} but this is just an optical effect due to the use of different scales in the vertical axis.}.

Fig.~\ref{fig:distances_with_correlation_Sigma_1400} shows that in most of the bins the mean values of the posteriors (black points) are closer to the truth than the ones of the priors (red points). Besides, even if the uncertainties in the priors are reduced with respect to $\Sigma=100$, they are still larger than the uncertainties in the posteriors. In particular, we see in the right panel (bottom row) that the distances of the posteriors in the bins corresponding to the peak of the top jet mass distribution are considerably closer than the priors in terms of significance illustrating how the inference process corrects the bias introduced by the priors. The posterior parameters result to be relatively precise but, as we have already discussed in Section \ref{sec:tight_priors}, they still miss the truth, and in some bins at more than 2$\sigma$, because the transfer matrix is approximate and the allowed exploration of parameter space is strongly constrained for tight priors. 

A remarkable example where the tight restriction on the priors leads the posterior to fail to hit the true value occurs in the fifth bins of the jet mass distributions (that is, for the mass in the interval [170,175] GeV). We see that the maximal separation between the prior and the truth in the top jet mass distribution takes place precisely in that bin. Since that separation is quite large in terms of the allowed variation of the prior, the way that the inference manages to deal with this difficulty is to compensate the insufficient growth of the top jet mass posterior by increasing the posterior of the QCD jet mass. This is in complete correspondence with Fig.~\ref{fig:inference_with/without_correlation_sigma_1400} and exemplifies, in addition, the correlations of the distances $D$ between classes. It is worth observing that the mentioned compensation is rather milder in the case of $\Sigma=100$ as is shown both in Fig.~\ref{fig:inference_with/without_correlation_sigma_100} and Fig.~\ref{fig:distances_with_correlation_Sigma_100}. Since we see from Figs.~\ref{fig:inference_with/without_correlation_sigma_100} and \ref{fig:inference_with/without_correlation_sigma_1400} that the uncertainties in the posteriors are generally larger for $\Sigma=100$ than for $\Sigma=1400$, this result suggests that it might be preferable to work with looser priors at the expense of having larger uncertainties in the posteriors, or in the other way around, posteriors corresponding to tighter priors may be relatively more precise but in values more distant from the truth.

\begin{figure}[h!]
    \begin{tabular}{cc}
    \includegraphics[width=7.2 cm]{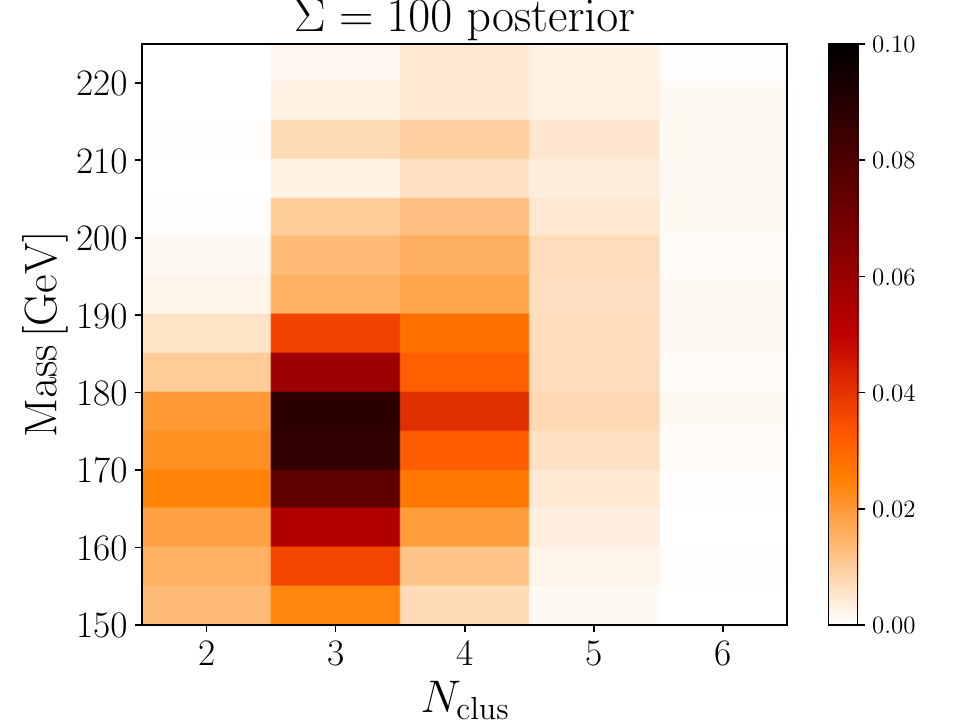} & 
    \includegraphics[width=7.2 cm]{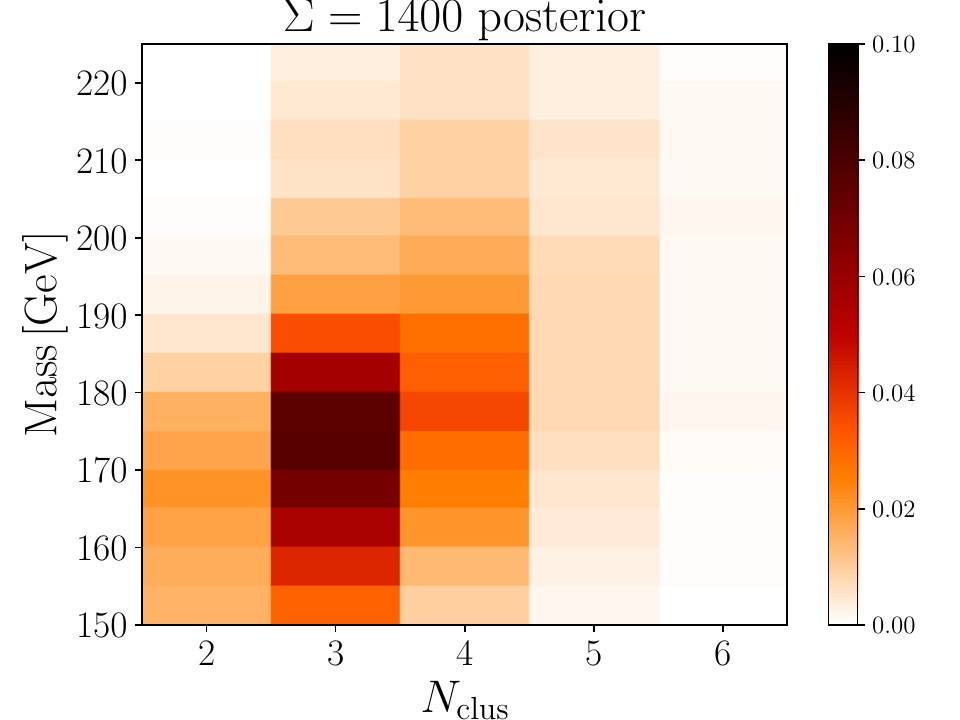}
    \end{tabular}
%    \vspace*{-2.cm}
    \begin{center}
    \includegraphics[width=7.2 cm]{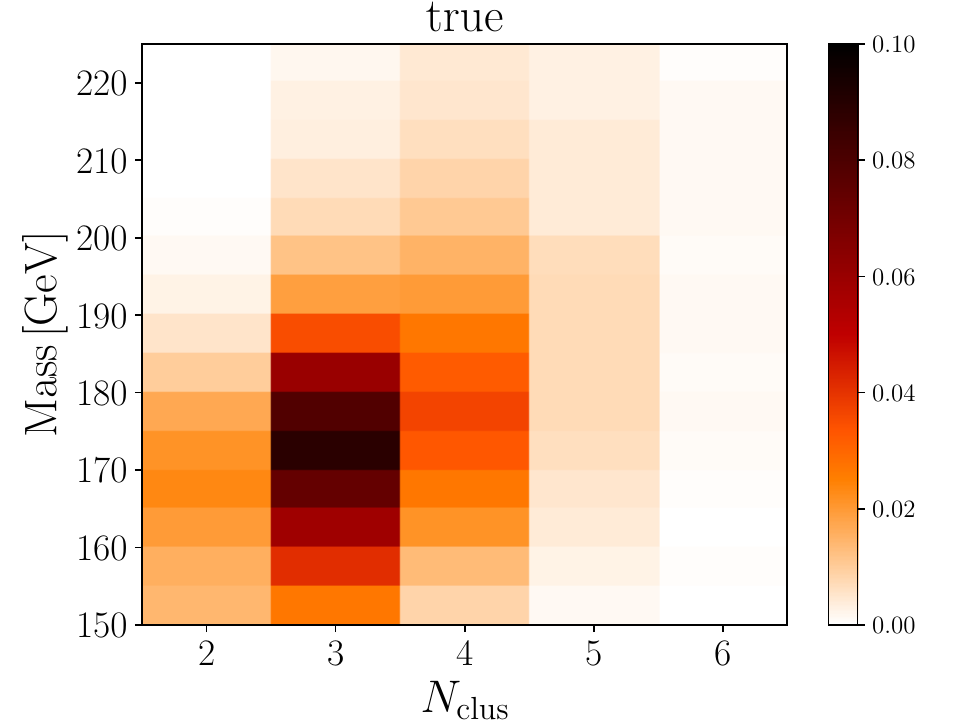} 
    \end{center}
%    \vspace*{-1.cm}
    \caption{MAP for the posterior $p(\vec x)$ of the model with correlations for the case of  loose priors (top left panel), tight priors (top right panel) compared to truth PDFs (bottom panel).}
    \label{fig:tagger_prob}
\end{figure}

Finally, we compare the outcome of the different frameworks in the full two-dimensional distributions on the characterizing observables $\vec{x}=\{N_{\mathrm{clus}},\mathrm{Mass}\}$. We show in Fig.~\ref{fig:tagger_prob} the Maximum a Posteriori (MAP) for the posterior $p(\vec x)$ of the model with correlations, for different cases.  To make a quantitative comparison of the similarity between the posteriors in different frameworks, the priors, and the true PDFs, it is suitable to use the Kullback–Leibler (KL) divergence \cite{Bishop}, which is a measure of the {\it statistical difference} between a model PDF and the true PDF.  We summarize all the relevant KL-divergence distances to their corresponding true PDF in Table~\ref{tab:KL}. In this table we compare KL-divergence distances between the prior distributions and true distributions (KL(prior,true) in the fourth column) to the distances between posterior and the true distributions (KL(posterior,true) in the fifth column) for the model with and without correlations (``yes'' and ``no'' in the third column of the table stand for the former and the latter, respectively). As before, we consider two scenarios for prior spread here: loose ($\Sigma=100$) and tight ($\Sigma=1400$). We calculate the distances separately for three categories of events: QCD events, top events and both QCD and top events.

It can be seen from Table~\ref{tab:KL} that:  1) the KL-divergence decreases if the correlations are included in the model compared to the model without correlations, 2) KL(posterior,true) < KL(prior,true) in the case of the model with correlations is taken into account, 3) KL-divergence is slightly smaller for loose priors compared to tight priors in the case of the model with correlations taken into account(which is in accordance with our earlier results), 4) KL-divergence is smaller for QCD events compared to top events (because the QCD priors are closer to the true values), 5) as expected, KL-divergence is usually smaller for the dataset that includes both QCD and top events. These results summarize the performance of our proposed objective of determining the underlying distributions in top events with correlation among the characterizing observables.

\begin{table}[h]
\centering
\begin{tabular}{|c|c|c|c|c|}
\hline
Class & Prior spread & Correlated model & KL(prior, true) & KL(posterior, true)\\
\hline\hline
QCD & tight & yes & 0.022 & {\bf 0.00175} \\
%\hline
Top & tight & yes & 0.038 & {\bf 0.011}\\
%\hline
both & tight & yes & 0.022 & {\bf 0.0108} \\
\hline
QCD & loose & yes & 0.022 & {\bf 0.00174} \\
%\hline
Top & loose & yes & 0.039 & {\bf 0.0096} \\
%\hline
both & loose & yes & 0.022 & {\bf 0.0105} \\
\hline
QCD & tight & no & 0.025 & 0.13\\
%\hline
Top & tight & no & 0.11 & 0.45 \\
%\hline
both & tight & no & 0.039 & 0.016 \\
\hline
QCD & loose & no & 0.024 & 0.16\\
%\hline
Top & loose & no & 0.12 & 0.48\\
%\hline
both & loose & no & 0.038 & 0.01 \\
\hline
\end{tabular}
\caption{Posterior to true and prior to true KL-divergence distances in the different frameworks.  We have boldfaced the results corresponding to learning each class true distributions when exploiting the correlations, which are among the main motivations for this work.}
\label{tab:KL}
\end{table}

\section{Discussion and conclusions}
\label{sec:conclusions}

In this manuscript, we have proposed a completely Bayesian strategy to study mixture models with correlated discrete variables, where the correlation can be learned from simulations, and we have applied said strategy to learn the density of QCD and top jets from measured jets at the LHC. We have defined a set of interesting variables which are either discrete or discretized and learned the relevant class-dependent multinomial distributions. We have shown how the quality of the model depends on the choice of prior and, more importantly, on the ability of the model to incorporate multidimensional information, which we do by estimating the per-class dependence with existing simulators.

We have shown how, although the model without correlations is a good model of the full data, the specific classes are better modeled when correlations are taken into account. We have proposed specific metrics that quantify the improvement of the posterior with respect to the prior and allow us to compare different models. In all cases, we have observed that the top distribution suffers from the largest impact of correlations and thus is the most improved when the model incorporates them. We speculate that the methodology introduced to mix data-driven one-dimensional estimators and simulator-driven correlations is useful beyond this specific example or application. In particular, physics analysis where the backgrounds are data-driven could use this approach to correct the inference on the data in the signal region with simulations. The usefulness of this will depend on how trustworthy the multidimensional data-driven template is, which needs to be assessed in a case-by-case basis. However, we highlight the fact that because correlations are defined implicitly through simulators, we can incorporate highly non-trivial dependencies.

In the future, this model could be extended by incorporating more classes (such as $W$ jets), and also by attempting to infer the correlations from the data itself. Of course, there is no free lunch and incorporating data-driven correlations should be compensated by additional model nuances.

\vspace*{-0.1cm}
\section{Acknowledgements}
This work was partially supported by CONICET. M.S. acknowledges the support in part by the DOE grant de-sc0011784 and NSF  OAC-2103889, OAC-2411215, and OAC-2417682.

\clearpage
\appendix
\section{Expectation-maximization algorithm}\label{app:em_algorithm}
In this appendix, we present in a schematic way the technical details of the expectation-maximization algorithm used to calculate the transfer matrices previously introduced in subsection \ref{sec:num_details}. The reader is referred to this section for a brief introduction to the EM algorithm and the explanation of the notation.

\begin{algorithm}
\caption{The EM algorithm for per-class transfer matrix estimation.}\label{alg:em_algorithm}
\begin{algorithmic}
\State $N^{\mathrm{MC},k}_{ij}\gets $Number of events in Monte Carlo sample for class $k$ in bin $(i,j)$.
\State $N^{\mathrm{MC},k}\gets \sum_{i}\sum_{j}N^{\mathrm{MC},k}_{ij}$
\State Initialize $C^{k,0}_{ij;i'j'}$
\State Initialize $\alpha^{k,0}_{i'}$
\State Initialize $\beta^{k,0}_{j'}$
\State $\mathcal{L}_{0} \gets \sum_{i=1}^{D_{1}}\sum_{j=1}^{D_{2}}N^{\mathrm{MC},k}_{ij}\ln \left(\sum_{i'=1}^{D_{1}}\sum_{j'=1}^{D_{2}} C^{k,0}_{ij;i'j'}\alpha^{k,0}_{i'}\beta^{k,0}_{j'}\right)$
\For{t=1,\dots,Max Epochs}
    \State \textbf{E-Step}
    %\\
    \State $\gamma^{k,t}_{ij;i'j'}=p_{t}(i'j'|ij)=\frac{p_{t}(ij,i'j')}{p_{t}(ij)}\gets \frac{C^{k,t-1}_{ij;i'j'}\alpha^{k,t-1}_{i'}\beta^{k,t-1}_{j'}}{\sum_{i'}\sum_{j'}C^{k,t-1}_{ij;i'j'}\alpha^{k,t-1}_{i'}\beta^{k,t-1}_{j'}}$
    
    \Comment{Probability of event in marginal bin $(i',j')$ if measured in bin $(i,j)$.}
    %\\
    \State \textbf{M-Step}
    %\\
    \State $N^{\mathrm{MC},k,t}_{i'j'} \gets \sum_{i}\sum_{j} N^{\mathrm{MC},k}_{ij}\gamma^{k,t}_{ij;i'j'}$
    
    \Comment{Estimated number of events in marginal bin $(i',j')$.}
    \State $N^{\mathrm{MC},k,t}_{i'} \gets \sum_{j'} N^{\mathrm{MC},k,t}_{i'j'}$
    \State $N^{\mathrm{MC},k,t}_{j'} \gets \sum_{i'} N^{\mathrm{MC},k,t}_{i'j'}$
    \\
    \State \textit{Update $\alpha$}
    %\State $f_{i'} = \sum_{i}\sum_{j'}\gets \sum_{i}\sum_{j}\sum_{j'}\frac{N^{\mathrm{MC,k}}_{ij}}{N^{\mathrm{MC,k}}}\gamma^{k,t}_{ij;i'j'}$
    %\State $\alpha^{k,t}_{i'} \gets \frac{f_{i'}}{\sum_{i'}f_{i'}}$
    \State $\alpha^{k,t}_{i'}=p_{t}(i')\gets \frac{N^{\mathrm{MC},k,t}_{i'}}{N^{\mathrm{MC},k}}$
    \\
    \State \textit{Update $\beta$}
    %\State $f_{j'} \gets \sum_{i}\sum_{j}\sum_{i'}\frac{N^{\mathrm{MC,k}}_{ij}}{N^{\mathrm{MC,k}}}\gamma^{k,t}_{ij;i'j'}$
    %\State $\beta^{k,t}_{j'} \gets \frac{f_{j'}}{\sum_{j'}f_{j'}}$
    \State $\beta^{k,t}_{j'}=p_{t}(j')\gets \frac{N^{\mathrm{MC},k,t}_{j'}}{N^{\mathrm{MC},k}}$
    \\
    \State \textit{Update $C$}
    %\State $f_{ij;i'j'} \gets N^{\mathrm{MC,k}}_{ij}\gamma^{k,t}_{ij;i'j'}$
    %\State $C^{k,t}_{ij;i'j'} \gets \frac{f_{ij;i'j'}}{\sum_{i}\sum_{j}f_{ij;i'j'}}$
    \State $C^{k,t}_{ij;i'j'}=p_{t}(ij|i'j')=\frac{p_{t}(ij,i'j')}{p_{t}(i'j')}=p_{t}(i'j'|i,j)\frac{p_{t}(i,j)}{p_{t}(i'j')} \gets \gamma^{k,t}_{ij;i'j'}\frac{N^{\mathrm{MC},k}_{ij}}{N^{\mathrm{MC},k,t}_{i'j'}}$
    \\
    \State \textit{Update likelihood value}
    \State $\mathcal{L}_{t}\gets \sum_{i=1}^{D_{1}}\sum_{j=1}^{D_{2}}N^{\mathrm{MC},k}_{ij}\ln \left(\sum_{i'=1}^{D_{1}}\sum_{j'=1}^{D_{2}} C^{k,t}_{ij;i'j'}\alpha^{k,t}_{i'}\beta^{k,t}_{j'}\right)$
    \\
\If{$\mathcal{L}_{t} \leq \mathcal{L}_{t-1}$}    
\State Break for loop
\EndIf
\EndFor
\State Return $C^{k,t}_{ij;i'j'}$
\end{algorithmic}
\end{algorithm}

\newpage

\bibliography{biblio_arxiv}

\end{document}